\documentclass[pre,superscriptaddress,twocolumn,showpacs]{revtex4}
\usepackage{epsfig,amsmath,amssymb,graphics,color,calc}

\newcommand{\be}{\begin{equation}}
\newcommand{\ee}{\end{equation}}
\newcommand{\Ncorr}{N_{\rm corr}}
\newcommand{\ba}{\begin{eqnarray}}
\newcommand{\ea}{\end{eqnarray}}

\begin{document}

\title{Spatial correlations in the dynamics of glassforming liquids: 
Experimental determination of their temperature dependence}

\author{C.~Dalle-Ferrier}
\affiliation{Laboratoire de Chimie Physique, UMR 8000,
Universit{\'e} Paris Sud and CNRS,
B{\^a}t. 349, 91405 Orsay, France}

\author{C. Thibierge}
\affiliation{Service de Physique de l'{\'E}tat Condens{\'e} 
(CNRS/MIPPU/URA 2464), DSM/DRECAM/SPEC,
CEA Saclay, P.C. 135, Gif sur Yvette, F-91191 Cedex, France}

\author{C. Alba-Simionesco}
\affiliation{Laboratoire de Chimie Physique, UMR 8000,
Universit{\'e} Paris Sud and CNRS,
B{\^a}t. 349, 91405 Orsay, France}

\author{L.~Berthier}
\affiliation{Joint Theory Institute, Argonne National Laboratory and
University of Chicago,
5640 S. Ellis Av., Chicago, Il 60637, USA}

\altaffiliation{Permanent address: Laboratoire des Collo{\"\i}des, Verres
et Nanomat{\'e}riaux, UMR 5587, Universit{\'e} Montpellier II and CNRS,
34095 Montpellier, France}

\author{G.~Biroli}
\affiliation{Service de Physique Th{{\'e}o}rique
Orme des Merisiers -- CEA Saclay, 91191 Gif sur Yvette Cedex, France}

\author{J.-P.~Bouchaud}
\affiliation{Service de Physique de l'{\'E}tat Condens{\'e} (CNRS/MIPPU/URA 2464), DSM/DRECAM/SPEC,
CEA Saclay, P.C. 135, Gif sur Yvette, F-91191 Cedex, France}
\affiliation{Science \& Finance, Capital Fund Management
6-8 Bd Haussmann, 75009 Paris, France}

\author{F. Ladieu}
\affiliation{Service de Physique de l'{\'E}tat Condens{\'e} (CNRS/MIPPU/URA 2464), DSM/DRECAM/SPEC,
CEA Saclay, P.C. 135, Gif sur Yvette, F-91191 Cedex, France}

\author{D. L'H{\^o}te}
\affiliation{Service de Physique de l'{\'E}tat Condens{\'e} (CNRS/MIPPU/URA 2464), DSM/DRECAM/SPEC,
CEA Saclay, P.C. 135, Gif sur Yvette, F-91191 Cedex, France}

\author{G. Tarjus}
\affiliation{LPTMC, UMR 7600, Universit{\'e} Pierre \& Marie Curie and CNRS,
4, Place Jussieu, 75252 Paris Cedex 05, France}

\date{\today}

\begin{abstract}
We use recently introduced three-point dynamic susceptibilities to obtain an experimental determination of the
temperature evolution of the number of molecules, $\Ncorr$, that are dynamically correlated during the structural relaxation of supercooled liquids. We first discuss in detail the physical content of  three-point functions that relate the sensitivity of the averaged two-time dynamics to external control parameters (such as temperature or density), as well as their connection to the more standard four-point dynamic susceptibility associated with dynamical heterogeneities. We then demonstrate that these functions can be experimentally determined with a good precision. We gather available data to obtain the temperature dependence of $\Ncorr$ for a large number of supercooled liquids over a wide range of relaxation timescales from the glass transition up to the onset of slow dynamics. We find that $\Ncorr$ systematically grows when approaching the glass transition. It does so in a modest manner close to the glass transition, which is consistent with an activation-based picture of the dynamics in glassforming materials. For 
higher temperatures, there appears to be a regime where $\Ncorr$ behaves as a power-law of the relaxation time. Finally, we find that the dynamic response to density, while being smaller than the dynamic 
response to temperature, behaves similarly, in agreement with theoretical expectations.   
\end{abstract}



\maketitle

\section{Introduction}

\label{intro:section}

One of the puzzling questions concerning the glass transition of liquids and polymers is the fact that the dramatic temperature dependence characterizing the slowing down of relaxation and viscous flow occurs with no significant variation of the typical structural length associated with static pair 
correlations (see however \cite{sid}). For instance, the static structure factor of a liquid barely changes from the high-temperature ``ordinary'' liquid down to the glass, while the main ($\alpha$) relaxation time may rise by up to fifteen orders of magnitude. Yet, on intuitive ground, it is tempting to ascribe the strong temperature dependence of the dynamics and the ubiquity of the phenomenon, irrespective of molecular details, to a collective or cooperative behavior characterized by a lengthscale that grows as one approaches the glass transition. This is indeed the core of most theories of the glass transition~\cite{DS,walter},
starting with the picture of ``cooperatively rearranging regions'' put forward by Adam and Gibbs in the mid-sixties~\cite{AG}.

From quite general arguments, one expects that a seemingly diverging, or at least long, timescale must be accompanied by a seemingly diverging, or at least large, lengthscale~\cite{MS0}.
Tracking such a lengthscale has been a long-standing goal of studies on glassforming materials. An exciting development in this direction has been the observation and description of the heterogeneous nature of the dynamics in viscous liquids and polymers~\cite{ediger,sillescu,richert}. Experiments carried out at temperatures near the glass transition and computer simulations on model liquids have shown that the dynamics of the molecules is indeed spatially correlated, with regions in space being characterized by mobilities or relaxation times substantially different from the average for a timescale at least equal to the main relaxation time. These dynamic heterogeneities come with (at least) one lengthscale that is typical of the correlations in space of the dynamics.

Measuring the lengthscale(s) associated with the spatial correlations of the dynamics is however a difficult task. Experimentally, the most direct measurements have been performed by means of solid-state multidimensional nuclear magnetic resonance, providing estimates of a few nanometers (between 5 and 10 molecular diameters)~\cite{ediger,sillescu,exp0,mark,encoremark}. Unfortunately, these measurements can only be done in a narrow temperature range close to the glass transition, thereby precluding any study of the temperature dependence. Other types of dynamic lengthscales have been extracted from experimental data, mostly from crossover phenomena (e.g., between Fickian diffusion and non-Fickian translational motion~\cite{wang}) or by some sort of dimensional analysis (e.g., from the product of a diffusion coefficient by a relaxation time or by the viscosity~\cite{fischer}; see also \cite{Donth} and the discussion below).

Computer simulations give the opportunity to study the properties of the dynamic heterogeneities in more detail, with the restriction however that only the moderately viscous regime, many orders of magnitude away from the glass transition, is accessible. Sizes characteristic of one form or another of space-time correlations (strings~\cite{string}, microstrings~\cite{microstring}, compact ``democratic'' clusters~\cite{democratic}, etc.) have been determined. On top of those studies, a more generic means of extracting a lengthscale associated with the dynamic heterogeneities, inspired by the mean-field theory of spin glasses~\cite{FP}, has been proposed and quite thoroughly investigated: the spontaneous fluctuations around the average dynamics give rise to a four-point time-dependent correlation function, $\chi_{4}(t)$, called dynamic susceptibility~\cite{footnote}; its magnitude defines a ``correlation volume'' that follows a nonmonotonic variation with time $t$ and displays a maximum for a timescale of the order of the $\alpha$-relaxation time~\cite{silvio2,glotzer,lacevic,berthier,TWBBB}. This correlation volume shows, in numerical simulations, a significant increase as temperature is decreased. At the present time, however, the dynamic susceptibility $\chi_{4}(t)$ cannot be experimentally measured in glassforming liquids and polymers, which prevents any overlap between simulations and experiments. This is because measuring $\chi_4(t)$ in principle amounts to resolving the dynamic behavior of molecules in both space and time, which has proven possible up to now only in systems such as colloidal suspensions~\cite{luca,lucabis}, granular
systems~\cite{dauchot,durian}, and foams~\cite{mayer}. However, a promising path is to measure the non-linear response of glass forming liquids, for example their non-linear dielectric constant, which contains an information similar to $\chi_4(t)$~\cite{BBPRB,TLL}. 

In spite of the progress made in the last decade, as briefly surveyed above, many important questions remain unanswered. To list a few: is there a unique lengthscale associated with the dynamic heterogeneities? Is the same lengthscale involved in the slowing down of the $\beta$ and $\alpha$-relaxation? What is the temperature dependence of the dynamic lengthscale(s) over the whole range from the ordinary liquid phase to the glass? Is there a difference of behavior between the ``strong'' and the ``fragile'' glassformers? What kind of underlying static length, if any, may drive the growing dynamic length(s) in liquids and polymers ?

The purpose of the present paper is not to answer, nor even to address, all of the above questions. We focus on the experimental determination of the
temperature dependence of a length (more precisely, a volume or a typical number of molecules, $\Ncorr$) that characterizes the space-time correlations in
glassforming liquids. We do so by building upon the recent proposal made by some of us~\cite{science,I,II}. We study the experimentally accessible three-point
time-dependent correlation functions defined as the response of a two-point
time-dependent correlator to a change of an external control parameter, such
as temperature, pressure, or density. In Ref.~\cite{science} it has been shown
that the three-point correlation functions (that we will equivalently call
``three-point dynamic susceptibilities'') can be used to provide a lower bound
to the dynamic susceptibility $\chi_{4}(t)$. More recent 
work ~\cite{I,II,BBMR} has in fact established that these three-point
susceptibilities contain {\it bona fide} information 
on the heterogeneous nature of the dynamics.
In the present article, (i) we
analyze dynamic data on a variety of glassforming liquids to obtain
three-point dynamic susceptibilities; (ii) we stress the arguments (and the
conditions) that allow one to extract a number of dynamically correlated
molecules $\Ncorr(t)$ (or a correlation volume); (iii) we study the temperature dependence of $\Ncorr$ for a timescale of the order of the $\alpha$-relaxation time $\tau_{\alpha}$ 
and discuss its possible connection 
with the ``fragility'' ($i.e.$, the degree of departure of $\tau_{\alpha}$ from an Arrhenius temperature dependence) of the glassformers; 
 (iv) we compare the relative contributions of temperature and density fluctuations to dynamical
correlations, (v) finally, we discuss the connection between the extent of spatial dynamic correlations and some aspects of phenomenological descriptions of the glass transition, such as the degree of ``cooperativity'' of the relaxation processes and the nonexponential character of the relaxation functions.


\section{Experimental access to dynamic spatial correlations}
\label{assessment:section}

We have already mentioned that the natural quantity characterizing the spatial fluctuations of the dynamics, namely the dynamic susceptibility $\chi_{4}(t)$, has not so far been experimentally accessible in glassforming liquids. However, three-point correlation functions can be obtained as the response of a two-point correlator, commonly determined in experimental studies of the dynamics in glassformers, to a change in an external control parameter. Many aspects of the problem have been studied in great detail in Refs.~\cite{science,I,II}. Here, we mainly focus on the interpretation of the three-point dynamic susceptibilities and on the extraction from it of a lengthscale,
or to the least of a volume or a number of molecules.

\subsection{Preliminary: A phenomenological description}

\label{preliminary}

Consider a dynamic process that proceeds through thermal activation and is characterized by a typical time, $\tau$, which follows an Arrhenius temperature dependence, $\tau \propto \exp({\Delta}/{k_{B}T})$, where $\Delta$ is the activation energy barrier, $k_{B}$ the Boltzmann's constant, and $T$ the temperature. An example is provided by the diffusion of vacancies in a solid. The activation barrier may be due to very local interactions, but in any case it is crossed because of energy fluctuations coming from the environment. A simple reasoning that allows one to derive a ``correlation volume'' associated with the activated dynamics is as follows. One just asks what is the volume $V^{*}$ that is necessary to generate typical energy fluctuations of the order of the energy barrier $\Delta$? Typical energy fluctuations in a volume $V$ are given by the square root of the variance of the total energy $E$ in the $NVT$ ensemble, $\left\langle \delta E^{2}\right\rangle_{V} =\rho k_{B}T^{2}c_{v}V$, where $c_{v}$ is the specific heat per particle in $k_{B}$ units and $\rho$ is the mean density.
 Equating $\Delta$ with $\sqrt{\left\langle \delta E^{2}\right\rangle_{V}}$ gives immediately
\begin{equation}
\label{eq1:equation}
\rho V^{*} = \frac{\Delta^{2}}{k_{B}T^{2}c_{v}}.
\end{equation}
If $c_{v}$ does not vary much or even goes to zero as $T$ goes to zero (as in
a solid), then $V^{*}$ increases at least like $T^{-2}$ as temperature is
lowered and ultimately diverges at $T=0$. This result looks intriguing, but
the ``correlation volume'' so defined appears more as the result of an
exercise in dimensional analysis than motivated by describing spatial
correlations in the dynamics. Actually, the nature of \textit{what is exactly correlated in the volume} $V^{*}$ is unclear (if not misleading).

Despite the obvious weakness of the approach, one may try to extend it to describe the slow, presumably thermally activated, dynamics in viscous liquids. There, however, the activation energy is not constant. The temperature dependence of the main ($\alpha$) relaxation time $\tau_{\alpha}$ is ``super-Arrhenius'' and one must choose an operational procedure to define an effective, temperature-dependent activation energy. One way is to define it through the relation $\tau_{\alpha} \propto \exp({\tilde \Delta(T)}/{k_{B}T})$, another possibility being to take the derivative of $\log \tau_{\alpha}$ with respect to $1/T$, namely $\Delta(T) = k_{B} \frac{\partial \log(\tau_{\alpha})}{\partial(1/T)}$. These two quantities are very different in fragile glassformers
with a pronounced super-Arrhenius behavior, since
$\Delta (T) = \tilde\Delta (T) -\partial \tilde\Delta(T)/\partial \log T$.
Anyhow, arbitrarily choosing at this stage the second definition and repeating the above argument leads to
\begin{equation}
\label{eq2:equation}
\rho V^{*}=\frac{k_{B}}{c_{v}} \left( 
\frac{\partial \log\tau_{\alpha}}{\partial \log T} \right)^2_V,
\end{equation}
which reduces to Eq.~(\ref{eq1:equation}) when the dependence of $\tau_{\alpha}$ is simply Arrhenius. This again results in a volume $V^{*}$ or a number of molecules $N^{*}=\rho V^{*}$ which increases 
faster than  $T^{-2}$ as $T$ decreases and approaches the glass transition. [This ``correlation volume'' is not to be confused with the
``activation volume'' sometimes introduced to characterize the
pressure dependence of the $\alpha$-relaxation and defined as
$T \partial \log \tau_\alpha /\partial P \vert_T$.] 
A variant of the above reasoning was in fact proposed and further investigated by Donth~\cite{Donth,encoredonth}, with the replacement of $(1/c_{v})$ by $\vert \Delta(1/c_{v})\vert \approx \Delta c_{v}/c_{v}^{2}$, where $\Delta c_{v}$ is the jump in the heat capacity at the glass transition. In Donth's interpretation, $V^{*}$ (or $N^{*}$) is taken, with no justification, as the size of the ``cooperatively rearranging regions'' introduced by Adam and Gibbs~\cite{AG}.

The same line of reasoning can be followed in the $NPT$ ensemble
with the activation energy replaced by an activation enthalpy, leading
to a correlation volume
\begin{equation}
\label{eq2NPT:equation}
\rho V^* = \frac{k_{B}}{c_p} \left( \frac{\partial \log \tau_\alpha}{\partial \log T} \right)^2_P.
\end{equation}

It seems clear that for this approach to have any true physical meaning, it needs to be put on a much firmer basis and insight should be provided about the nature of the dynamic correlation. 
This is what we address below.

\subsection{Multi-point dynamic correlation functions and susceptibilities}

As was shown in Ref.~\cite{science} and further detailed in
Refs.~\cite{I,II}, multi-point space-time correlations in glassformers may be experimentally accessible by studying the response of the dynamics to a change in a control parameter, \textit{i.e.}, by inducing fluctuations in some conjugate quantity such as energy or density and monitoring its effect on the dynamics of the system. To provide a self-contained presentation, we briefly recall here the main arguments.

Standard experimental probes of the dynamics in liquids give access to the time-dependent auto-correlation function of the spontaneous fluctuations of some observable $O(t)$, $F(t)= \left\langle \delta O(0) \delta O(t)\right\rangle $, where $\delta O(t)=O(t)-\left\langle O\right\rangle$ represents the instantaneous value of the fluctuations of $O(t)$ from its ensemble
average $\langle O \rangle$.
One can think of $F(t)$ as being the average of a two-point quantity, $C(0,t)=\delta O(0) \delta O(t)$, characterizing the dynamics. This quantity fluctuates around its mean value, and information on the amplitude of those fluctuations is provided by the variance $\left\langle \delta C(0,t)^{2}\right\rangle$, where $\delta C(0,t)=C(0,t)-F(t)$. This allows to interpret $\chi_4(t)$ as a precise measure of dynamic
fluctuations  or heterogeneities, since $N\left\langle  \delta C(0,t)^{2}\right\rangle=\chi_{4}(t)$, where $N$ is the total number of particles in the system.
The associated spatial correlations show up more clearly when considering a ``local'' probe of the dynamics, like for instance an orientational correlation function measured by dielectric or light scattering experiments, which can be expressed as
\begin{equation}
\label{eq3:equation}
C(0,t) = \frac{1}{V}\int d^{3}r \, c(\mathbf{r};0,t),
\end{equation}
where $V$ is the volume of the sample and $c(\mathbf{r};0,t)$ characterizes the dynamics between times $0$ and $t$ around point $\mathbf{r}$. For example, in the above mentioned case of orientational correlations, $c(\mathbf{r};0,t)\propto \frac{V}{N}\sum_{i,j=1}^{N}\delta(\mathbf{r}-\mathbf{r}_{i})Y(\Omega_{i}(0))Y(\Omega_{j}(t))$, where $\Omega_{i}$ denotes the angles describing the orientation of molecule $i$, $\mathbf{r}_{i}(0)$  is the position of that molecule at time $0$, and
$Y(\Omega)$ is some appropriate rotation matrix element; the ``locality'' of the probe comes from the fact that it is dominated by the self-term involving the same molecule at times $0$ and $t$ or by the contribution coming from neighboring molecules.

The dynamic susceptibility can thus be rewritten as
\begin{equation}
\label{eq4:equation}
\chi_{4}(t) = \rho \int d^{3}r \left\langle \delta c(\mathbf{0};0,t)\delta c(\mathbf{r};0,t)\right\rangle,
\end{equation}
where the statistical translational invariance of the liquid has been taken into account and $\rho=N/V$ denotes the mean density. The above equation shows that $\chi_{4}(t)$ measures the extent of spatial correlation between dynamical events between times $0$ and $t$ at different points of the system, \textit{i.e.}, the spatial extent of the dynamic heterogeneities over a time span $t$.
Thanks to a number of recent theoretical and numerical studies~\cite{silvio2,glotzer,lacevic,berthier,TWBBB}, this point is now well documented.

Unfortunately, $\chi_{4}(t)$ has not so far been measurable in glassforming liquids and polymers. Furthermore, its physical interpretation is obscured by the fact that $\chi_4$ turns out to depend strongly 
on microscopic dynamics and thermodynamical ensemble (see \cite{I,II}). A related and experimentally feasible route to space-time correlations in glassformers consists in monitoring the response of the average dynamics to an infinitesimal change, say, of temperature. In the case of Newtonian dynamics, such as the molecules in a liquid, one can show that this response is given through a fluctuation-dissipation relation by a cross-correlation function involving the spontaneous fluctuations of the dynamics and those of the energy~\cite{science}. For simplicity, let us first focus on the $NVT$ ensemble. In this case one finds~\cite{science}:
\begin{equation}
\label{eq5:equation}
\chi_{T}^{NVT}(t) = \frac{\partial F(t)}{\partial T} \bigg\vert_{N,V}=\frac{N}{k_{B}T^{2}}\left\langle \delta E(0)\delta C(0,t)\right\rangle_{NVT},
\end{equation}
where $\delta E(0)=E(0)-\left\langle E\right\rangle$ is the fluctuation of the energy \textit{per molecule} at time $t=0$. The subscript $NVT$ means that the statistical average is performed in the $NVT$ ensemble. Expressing the fluctuations in terms of local quantities (which is always possible~\cite{I,II,hansen})
as before leads to
\begin{equation}
\label{eq6:equation}
\chi_{T}^{NVT}(t) = \frac{\rho}{k_{B}T^{2}}\int d^{3}r \left\langle \delta e(\mathbf{0};0)\delta c(\mathbf{r};0,t)\right\rangle_{NVT},
\end{equation}
where we have introduced $e(\mathbf{r};t)$ as the energy density per molecule (hence, $N E(t)=\rho \int d^{3}r \, e(\mathbf{r};t)$). Note that the correlation functions in Eqs.~(\ref{eq5:equation}, \ref{eq6:equation}) are expected to be negative since an increase in energy is likely to accelerate the dynamics, producing a negative change in $C(0,t)$.

The above equation shows that the three-point dynamic susceptibility $\chi_{T}(t)$, up to a factor $\rho/(k_{B}T^{2})$, measures the spatial extent of the correlation between a fluctuation of energy at time $0$ and at some point $\mathbf{0}$, and the change in the dynamics occurring at another point $\mathbf{r}$ between times $0$ and $t$. Clearly, $\chi_{T}(t)$ qualifies as a relevant
quantitative indicator of space-time correlations in glassformers, whose temperature (or density) dependence is now experimentally accessible.

How does this alternative indicator relate to the dynamic susceptibility $\chi_{4}(t)$? On physical ground, it seems reasonable that energy fluctuations provide the main source of fluctuations in the dynamics, especially within a picture of slow activated relaxation with sizeable energy barriers (recall that we consider the $NVT$ ensemble). As a result, spatial correlations between changes in the dynamics at different points can be mediated by the energy fluctuations, as thoroughly discussed in Ref.~\cite{I}. This view is supported by an exact relation that can be derived between $\chi_{T}(t)$ and $\chi_{4}(t)$~\cite{science,I,II}.
From the Cauchy-Schwarz inequality, $\left\langle \delta E(0)\delta C(0,t)\right\rangle_{NVT}^{2}\leqslant \left\langle \delta E(0)^{2}\right\rangle_{NVT} \left\langle \delta C(0,t)^{2}\right\rangle_{NVT}$, and the thermodynamic relation, $N\left\langle \delta E(0)^{2}\right\rangle_{NVT}=k_{B}T^{2}c_{v}$, with $c_{v}$ the constant volume specific heat per molecule in $k_{B}$ units, one indeed obtains
\begin{equation}
\label{eq7:equation}
\chi_{4}^{NVT}(t) \geq \frac{k_{B}T^{2}}{c_{v}} \left[ \chi_{T}^{NVT}(t) 
\right]^{2},
\end{equation}
so that the experimentally accessible response $\chi_{T}(t)$ can be used to give a lower bound on the well-studied susceptibility $\chi_{4}(t)$. A rapidly growing energy-dynamics correlation as temperature decreases therefore directly implies a rapidly growing dynamics-dynamics correlation~\cite{I}.

Actually, it can be shown that $\chi_{4}$ and $\chi_{T}$ are more precisely related by
\begin{equation}
\label{eq8:equation}
\chi_{4}^{NVT}(t) = \frac{k_{B}T^{2}}{c_{v}}
\left[ \chi_{T}^{NVT}(t) \right]^{2}+\chi_{4}^{NVE}(t).
\end{equation}
The dynamic susceptibility in the $NVE$ ensemble, $\chi_{4}^{NVE}(t)$, is a variance and therefore always positive, which of course is compatible with the inequality in Eq.~(\ref{eq7:equation}).
It is no more accessible to experimental measurements than $\chi_{4}^{NVT}(t)$, but it can nevertheless be computed in numerical simulations and in some theoretical approaches, as
discussed in Refs.~\cite{I,II}. In particular, it was shown from simulations on two different models of glassforming liquids that for times of the order of $\tau_{\alpha}$ (and in the range covered by the simulations) $\chi_{4}^{NVE}$ becomes small compared to $(k_{B}T^{2}/c_{v})(\chi_{T}^{NVT})^{2}$ when temperature is low enough and the dynamics becomes truly sluggish, which then makes the lower bound in Eq.~(\ref{eq7:equation}) a good estimate of $\chi_{4}^{NVT}$. More importantly, the temperature dependences of $\chi_{4}^{NVE}$ and $\vert \chi_{T}^{NVT}\vert$ for $t\sim\tau_{\alpha}$ are found 
to be very similar, as can be proved exactly within the framework of the Mode-Coupling Theory. Furthermore, for Brownian dynamics $\chi_4^{B}$ behaves as $\chi_{4}^{NVE}$. These results
strongly support the claim that the dynamic spatial correlations in glassforming liquids are in fact fundamentally contained in the three-point dynamic susceptibility $\chi_{T}$ (see ~\cite{science,I,II}
and the discussion in Sec.~\ref{jp-discussion} below). In this respect, it is satisfying that $\chi_T(t = \tau_\alpha)$ is found to be independent, for large $\tau_\alpha$, of the type of dynamics (Newtonian, Brownian or Monte-Carlo), as is the very phenomenon of glassy slowing down~\cite{gleim,szamel,LJMC,BKSMC}.  

\subsection{Experimental conditions: $NPT$ thermodynamic ensemble}

\label{2C}

Most experimental situations in glassforming liquids and polymers correspond to the $NPT$ ensemble. In order to carry out the analysis of the experimental data, one must first discuss the modifications brought about in the above presentation by switching the statistical ensemble from $NVT$ to $NPT$.

The counterparts of Eqs.~(\ref{eq5:equation}-\ref{eq8:equation})
are simply obtained by replacing energy by enthalpy and constant volume specific heat by constant pressure specific heat. For instance, one now has
in place of Eq.~(\ref{eq6:equation})
\be
\chi_T^{NPT}(t) = \frac{\rho}{k_{} T^2} \int d^3 r \langle \delta h
({\bf 0};0) \delta c ({\bf r}; 0,t)\rangle_{NPT},
\label{eq:chiNPT}
\ee
 where
$\chi_T^{NPT}(t)$ is the derivative of $F(t)$
with respect to $T$ when $N$ and $P$ are kept constant and where
$h({\bf r},t)$ is the enthalpy density per molecule.
Equation~(\ref{eq8:equation}) is similarly changed into
\begin{equation}
\label{eq13:equation}
\chi_{4}^{NPT} = \frac{k_{B}T^{2}}{c_{p}}(\chi_{T}^{NPT})^{2}
+\chi_{4}^{NPH},
\end{equation}
where the first term in the right-hand side is a lower bound for the nonlinear susceptibility $\chi_{4}^{NPT}$ and $\chi_{T}^{NPT}$ is, up to a factor $\rho/(k_{B}T^{2})$, the integral of a correlation function between local fluctuations of the dynamics and of the enthalpy.

Under constant pressure conditions, a change in temperature is accompanied by a change in density. As a result, one can separate in the response of $F(t)$ to a change of temperature at constant pressure the effect of temperature at constant density, described by $\chi_{T}^{NVT}$ defined in Eq.~(\ref{eq5:equation}), and the effect of density at constant temperature, characterized by $\chi_{\rho}^{NPT}(t) = \partial F(t)/\partial \rho \vert_{T}$.
This allows us to generalize Eq.~(\ref{eq13:equation}) to
\begin{equation}
\label{eq14:equation}
\chi_{4}^{NPT} = \frac{k_{B}T^{2}}{c_{v}}(\chi_{T}^{NVT})^{2}+\rho^{3}k_{B}T\kappa_{T}(\chi_{\rho}^{NPT})^{2}+\chi_{4}^{NVE},
\end{equation}
where $\kappa_{T}$ is the isothermal compressibility of the liquid. The last term of the right-hand side being positive, one can now use the sum of the two first terms as a 
lower bound to $\chi_{4}^{NPT}$, which is slightly different from that considered above, but which explicitly takes into account all sources of fluctuations.

In any case, since both $\chi_{T}^{NVT}$ and $\chi_{\rho}^{NPT}$ can be
experimentally obtained, the above expressions will allow us to evaluate the relative magnitudes of the contribution due to energy fluctuations and to density fluctuations in the space-time correlations, which represents an interesting issue {\it per se}.

\subsection{From dynamic susceptibilities to correlation volumes}

\label{from}

So far, we have presented a rigorous framework in which the spatially heterogeneous nature of the dynamics in glassforming materials is described through susceptibilities that involve spatial correlations associated with the dynamics. One expects that those correlations die out at large separation, so that their integral over space provides a measure of the characteristic ``correlation volume'' at a given time $t$. It is then tempting, in light of Eqs.~(\ref{eq4:equation}, \ref{eq6:equation}, \ref{eq:chiNPT}), to interpret an increase with decreasing temperature of either $\chi_{4}(t)$ or $k_{B}T^{2}\vert  \chi_{T}(t)\vert$ (for $t$ of the order of $\tau_{\alpha}$) as a signature of a dynamic correlation lengthscale or correlation volume that grow as the glass transition is approached.

Although the above considerations appear most plausible and indeed, as will be discussed below, are supported by a series of numerical and theoretical arguments, it is also worth stressing what could go wrong in the corresponding line of thought. With $\chi_{4}(t)$ and $\chi_{T}(t)$ one only has access to the integral over space of a correlation function, and two potential difficulties must be addressed.
First, the correlation function should go to zero at large distances. Although this may sound as a trivial requirement,  it is more subtle than anticipated and it depends on the statistical ensemble under consideration~\cite{I}. It is a well-known feature, even in the simplest case of the static pair correlation~\cite{hansen}, that large distance limits of correlation functions in different ensembles differ by terms of the order $1/N$ or $1/V$. The difference is negligible in the thermodynamic limit, except when the function is integrated over the whole volume, as it is in susceptibilities~\cite{lebo}. It turns out that this is not an issue when all the conserved quantities are let free to fluctuate. For one-component molecular liquids, this corresponds to the $NPT$ ensemble in which both energy and density may fluctuate; on the other hand, the (negative) asymptotic background contribution has to be taken into account in the $NVT$ ensemble, which we will therefore no further consider.

Secondly, in the absence of more direct information on the spatial dependence of the correlation functions, extracting from their integral over space a ``volume of correlation'' (the derivation of a ``correlation length'' will be discussed below) requires an estimate of the amplitude of the correlation function to properly normalize the function. Indeed, an increase of $\chi_{4}$ or $k_{B}T^{2}\vert  \chi_{T}\vert$ as $T$ decreases could be due not only to the \textit{extension in space} of the associated correlations, the phenomenon one is looking for, but also to a variation of the typical \textit{amplitude} of the correlations. Disentangling the two effects is thus a prerequisite.

In the following, we focus on a timescale of the order of the $\alpha$-relaxation time $\tau_{\alpha}$, for which the susceptibilities $\chi_{4}$ and $\vert  \chi_{T}\vert$ are found to reach a maximum. Relaxation on this timescale, at least at low temperature, is known to be due to rare dynamic events associated with a significant apparent activation energy barrier (compared to $k_{B}T$). This implies that the relevant dynamical events typically affect $c({\bf r},0,t\sim \tau_{\alpha})$ on the order of a fraction of the mean value of the correlation function.
Therefore, provided the function $C(0,t)$ is normalized by its mean
value at $t=0$, $F(0)=\left\langle C(0,0)\right\rangle$, which we shall assume
from now on, such changes should be of order one and should not vary much with temperature.

The case of the dynamic susceptibility $\chi_{4}(t)$ is well documented. One expects that the typical magnitude of the auto-correlation function $\left\langle \delta c(\mathbf{0};0,t\sim \tau_{\alpha})\delta c(\mathbf{r};0,t\sim \tau_{\alpha})\right\rangle$ when $\mathbf{r}$ is in the vicinity of point $\mathbf{0}$ is of order one when the dynamics is dominated by activated, intermittent processes. Hence $\chi_{4}(t\sim \tau_{\alpha})$ is, up to a number of order unity, a measure of a correlation volume, or more properly (see the density factor in Eq.~(\ref{eq4:equation})) of a number of molecules that are dynamically correlated over a time span of the order of $\tau_{\alpha}$. More precisely, one can define
\begin{equation}
\label{eq9:equation}
N_{{\rm corr},4}= \max_{t}\lbrace\chi_{4}(t)\rbrace,
\end{equation}
the maximum occurring for $t \sim \tau_{\alpha}$.

Applying the same line of argument to the three-point susceptibility $\chi_{T}(t \sim \tau_{\alpha})$ is a little more tricky because one now faces a cross-correlation function between enthalpy fluctuation at time $0$ at a given point in space and fluctuation of the dynamics between $t=0$ and $t\sim \tau_{\alpha}$ in another place.

First, the local enthalpy fluctuations that influence a dynamic event must have time to travel to the vicinity of the ``active spot''. For instance, in the hydrodynamic limit, long-wavelength and small-amplitude enthalpy fluctuations follow a diffusive motion in a liquid, with a diffusion coefficient $D_{T}$; the range of influence of such fluctuations must therefore be less than a lengthscale of the order of $\sqrt{D_{T}\tau_{\alpha}}$. However, because $\tau_{\alpha}$ dramatically increases with decreasing $T$ whereas $D_{T}$ has a much weaker dependence, this hydrodynamic lengthscale is very large in viscous liquids and the upper bound it represents for dynamic correlations is of no real significance.

Secondly, the magnitude of the enthalpy fluctuations is temperature dependent, as shown by the thermodynamic relation $N\left\langle \delta H^{2}\right\rangle = k_{B}T^{2}c_{p}$. Removing this unwanted $T$-dependence requires to somehow normalize the  variance of the enthalpy, e.g., to define a local dimensionless fluctuation of enthalpy density, $\delta \hat{h}({\bf r})$, via $N\delta H= \rho \sqrt{k_{B}T^{2}c_{p}}\int d^{3}r \delta \hat{h}(\mathbf{r})$. However, one may imagine that activated dynamic events associated with the $\alpha$ relaxation and leading to a change in $\delta c$ of order one involve enthalpy fluctuations more sizeable than average, but \textit{rare}. An appealing alternative is then to introduce another dimensionless enthalpy density fluctuation, 
$\delta \bar{h}(\mathbf{r})$, through  
\be 
N\delta H\vert_{\alpha} = \rho \sqrt{k_{B}T^{2}\Delta c_{p}}\int d^{3}r \delta \bar{h}(\mathbf{r}), 
\ee
where $\Delta c_{p}$ is the ``configurational'' part of the heat capacity of a glassforming liquid, \textit{i.e.}, that in excess of the associated glass: this describes the contribution of those enthalpy fluctuations (indicated by the subscript $\alpha$ 
and whose typical size is given by
$N\langle \delta H\vert_{\alpha}^2 \rangle = k_{B} T^2 \Delta c_p$) 
that precisely disappear when the $\alpha$ relaxation is frozen out on the experimental timescale. Presumably, the remaining fast components of the enthalpy fluctuations are then unable to induce the $\alpha$ relaxation, or at least are very weakly correlated with the $\alpha$ relaxation processes. (Nonetheless, and as usual, the temperature dependence of $\Delta c_{p}$ can only be approximately determined experimentally by subtracting from the heat capacity of the liquid that of the crystal at the same temperature.)

Note that this distinction between typical and rare fluctuations is especially important in the presence of localized defects. Such defects, whose motion do not follow the standard hydrodynamic diffusion but is rather  akin to that of ``solitons'' conserving their energy (or, at constant pressure, their enthalpy), have a negligible influence on the thermodynamics, hence on $c_{p}(T)$, provided they are dilute enough. In the case where such dilute defects are responsible for the slow relaxation of the system, it is clear that an estimate of the dynamic correlation volume based on characterizing the magnitude of the relevant enthalpy fluctuations by means of $c_{p}$ would be meaningless, whereas $\Delta c_{p}$ in this instance would precisely capture the defect contribution and lead to a reasonable estimate, see
Ref.~\cite{II} for a discussion of defect models in this context.
For glassforming liquids in general, the difference between $c_{p}$ and $\Delta c_{p}$ is not as dramatic, and in most cases
remains of order of $c_{p}$ itself~\cite{BBT}.

From the above discussion, we conclude that an estimate of the number of molecules whose dynamics on the timescale of the $\alpha$ relaxation is correlated to a local enthalpy fluctuation is given by
\ba
\label{eq10:equation}
N_{{\rm corr},T} & \sim & \rho \left\vert \int d^{3}r \left\langle \delta \bar{h}(\mathbf{0};0)\delta c(\mathbf{r};0,t\sim \tau_{\alpha})\right\rangle \right\vert \nonumber \\
& = & \sqrt{\frac{k_{B}T^{2}}{\Delta c_{p}}} \max_{t}
\left\{  \vert \chi_{T}(t)\vert \right\},
\ea
where $\delta \bar{h}$ is the dimensionless fluctuation of energy density introduced above and where we have used the fact that $\vert \chi_{T}(t)\vert$ is maximum for $t\sim \tau_{\alpha}$~\cite{I,II}.
This is the central formula which bolsters our following analysis of experimental data in terms of number of dynamically correlated molecules.

\subsection{Discussion}

\label{jp-discussion}

Before moving on to the actual analysis of experimental data, we 
would like to stress a number of points.

{\it (i) $N_{{\rm corr},4}$ versus $N_{{\rm corr},T}$}---A priori, $N_{{\rm corr},4}$ and $N_{{\rm corr},T}$ need not coincide, nor should they share the same temperature dependence. The former is the typical number of molecules whose dynamics are correlated ``among themselves'', the latter the typical number of molecules whose dynamics are correlated ``to a local fluctuation of enthalpy''. For Newtonian dynamics, and if the bound in Eq.~(\ref{eq7:equation}) is nearly saturated, then these two numbers are actually quite different since:
\begin{equation}
\label{eq11:equation}
N_{{\rm corr},4}\sim N_{{\rm corr},T}^{2}.
\end{equation}
This is indeed the result found within Mode-Coupling Theory near its dynamic singularity~\cite{II}.
This difference is however entirely due to the presence of one (or several) conserved variables. The 
primary mechanism creating dynamic correlations is captured by $N_{{\rm corr},T}$ and is due to
the fact that any local perturbation coupled to the slow dynamics (energy, density, composition, etc.) 
affects the dynamics over a certain volume, which grows as the glass transition is approached. This
is measured by a certain dynamic response $\chi_d(|{\bf r}- {\bf r'}|,t)= \langle \partial c({\bf r},t)/\partial e({\bf r'},0)\rangle$ \cite{BBMR,I,II}. 
Now, summing over ${\bf r}$ for a given point perturbation at ${\bf r'}$ to define a correlation volume is {\it identical} to fixing ${\bf r}$ but 
perturbing the system uniformly in space (while staying in a linear regime) \cite{BBMR}. 
This is the fundamental reason why $\chi_T$, simply obtained by taking the derivative of $C$ with respect to
temperature, is related to a dynamical response.  By the same token, the  derivative of $C$ with respect to any 
quantity coupled to dynamics is also a dynamic response; in the regime where these quantities become large, we 
expect that all of them ($\chi_T, \chi_\rho, ...$) should behave similarly as a function of time and temperature, but 
with different prefactors (see \cite{I} for a detailed proof of this point). 
We shall check this prediction on experimental data in Sec.~V below.

Following the above reasoning, the difference between $N_{{\rm corr},T}$ and $N_{{\rm corr},4}$ comes from the
fact that if energy is conserved, a local energy fluctuation at ${\bf r}$ affects the dynamics in the surrounding, but
also propagates in space to create another {\it correlated} dynamical fluctuation elsewhere. This, in effect, transforms 
$N_{{\rm corr},T}$ into $N_{{\rm corr},T}^2$. This justifies that we will mostly focus below on $N_{{\rm corr},T}$ or $N_{{\rm corr},\rho}$ and 
less on $N_{{\rm corr},4}$. One should nevertheless keep in mind that 
some physical observables, such as the non-linear (cubic) response to 
an external
field, are directly related to $N_{{\rm corr},4}$ \cite{BBPRB}, and that 
both three-point and four-point susceptibilities
are deeply connected~\cite{science}.

{\it (ii) Prefactors}---The uncertainty concerning the prefactors involved in extracting correlation volumes or numbers of correlated molecules, in particular the normalization problem discussed above, or the difference between using $c_{p}$ and $\Delta c_{p}$, precludes a precise determination of the absolute value of those quantities. For instance, in the case of strong glassformers, the ratio $\Delta c_{p}/c_{p}$ may vary significantly, say from $1/2$ to $1/10$, between a fragile and a strong liquid. Choosing $c_{p}$ or $\Delta c_{p}$, or even $\Delta c_{p}/c_{p}^{2}$ as proposed by Donth~\cite{encoredonth}, then affects the relative estimates of the strong versus fragile glassformers. As found in Ref.~\cite{science}, the order of magnitude of the resulting correlation volumes appears nonetheless physically sensible. Here, we shall focus on the variation with temperature which, as shown in the next section, is less sensitive to the estimate of the prefactors.

{\it (iii) Equation~(\ref{eq2NPT:equation}) recovered}---The simple phenomenological analysis described in Sec.~\ref{preliminary} is recovered by assuming that the shape of the relaxation function $F(t)$ in the $\alpha$-relaxation regime does not vary with temperature, what is commonly referred to as ``time-temperature superposition''. The only temperature dependence of $F(t)$ then comes through that of the relaxation time $\tau_{\alpha}$, $F(t,T)=\Phi(t/\tau_{\alpha}(T))$ with $\Phi(0)=1$, which leads in the $NPT$ ensemble to
\begin{equation}
\label{eq12:equation}
N_{{\rm corr},T} \sim \sqrt{\frac{k_{B}}{\Delta c_{p}}} \left\vert
\Phi'(1) \frac{\partial \log \tau_{\alpha}}{\partial \log T} \right\vert,
\end{equation}
where the prime on $\Phi$ indicates a derivative. By using now Eq.~(\ref{eq11:equation}) and dropping the factor $\vert \Phi'(1)\vert$ that quantifies the stretching of the relaxation function and is thus of order one, one arrives at $N_{{\rm corr},4} \sim \frac{k_{B}}{\Delta c_{p}}({\partial \log \tau_{\alpha}}/{\partial \log T})^{2}$,
which is similar to Eq.~(\ref{eq2NPT:equation}).

The phenomenological derivation can thus be recast in a rigorous framework, which allows one to discuss the approximations made and, quite importantly, to understand the nature of the correlation that is probed. Essentially, the latter is the spatial correlation between local fluctuations of the dynamics and of the enthalpy. Any interpretation in terms of Adam-Gibbs-like cooperativity regions~\cite{Donth} requires additional, speculative steps that seem
of dubious validity, as discussed further in Sec.~\ref{conclusion}.

{\it (iv) Link with numerical simulations}---A major advantage of having a rigorous formulation of the space-time correlations in liquids, relating dynamic susceptibilities to multi-point correlation functions, is that the heuristic considerations that we have developed in this
and the previous subsections  can be checked on models of glassforming liquids. Computer simulations are extremely useful in this context: one is guaranteed that the features of the dynamic heterogeneities that are probed in experiments are indeed the same as in simulations, except that the latter allow a much more detailed analysis, in particular concerning the full spatial dependence of the multi-point correlation functions, and not only their integral over space such as $\chi_T$ or $\chi_4$. The numerical investigations so far carried out unambiguously confirm the interpretation of Eqs.~(\ref{eq10:equation}, \ref{eq11:equation}) for estimating the spatial extent of the correlations in the dynamics of glassforming liquids~\cite{I,II}.

{\it (v) Lengthscales}---The growth of the number of correlated molecules, $N_{{\rm corr},T}$, as one lowers the temperature, or equivalently of the associated correlation volume obtained by dividing $N_{{\rm corr},T}$ by the density, is a strong indication of a growing correlation length. However, the precise relation is not straightforward. There is no reason indeed for the correlation volume(s) to be compact and therefore to scale as $\Ncorr \sim \xi^3$. Actually, if such a behavior were true for $N_{{\rm corr},T}$, Eq.~(\ref{eq11:equation}) would lead to $N_{{\rm corr},4} \sim \xi^{6}$!
Extracting a lengthscale is thus a highly nontrivial task which requires additional information on the spatial dependence of the multi-point correlation functions involved in $\chi_{T}(t)$ and $\chi_{4}(t)$. Again, such information is provided by numerical simulations and by model theoretical calculations (e.g., in the framework of the mode-coupling approach, where $N_{{\rm corr},T} \sim \xi^4$ in the $\alpha$ regime, or in kinetically constrained models). The main result of both theoretical models and numerical simulations is that a {\it unique lengthscale} appears to characterize both $\chi_{T}(t)$ and $\chi_{4}(t)$, the associated correlation volumes being however noncompact~\cite{I,II,BBMR}.

\section{Data treatment and calibration}
\label{data}

We have shown in the previous section that an estimate of the number of dynamically correlated molecules, $N_{{\rm corr},T}$ (and via the lower bound, $N_{{\rm corr},4}$), can be obtained by measuring the sensitivity of two-time correlators to temperature changes, see Eq.~(\ref{eq10:equation}). A standard tool to access such a two-time correlator in supercooled liquids is dielectric spectroscopy, and we have gathered published (and original) dielectric data on a range of glassforming liquids. In addition, in a few cases, we have also considered relaxation data obtained by other techniques, photon correlation spectroscopy, dynamic light scattering, optical Kerr effect, and neutron scattering. In this section, we review our operational protocol for treating the experimental data and we examine the uncertainties due to systematic or statistical errors. Readers interested only in the main physical results may skip this section and 
focus directly on the following ones. 

\subsection{Fitting the relaxation data}
\label{fitting}

The quantity measured in dielectric spectroscopy is the linear susceptibility $\epsilon(\omega,T)$, which we have  described at each available temperature by an efficient empirical parametrization, the Havriliak-Negami (HN) form~\cite{refHavriliakNegami} or the one proposed by Blochowicz \textit{et al.}~\cite{refBlochowicz2006and2003} on the basis of the extended generalized Gamma function (GGE). The temperature dependence of the HN or GGE adjustable parameters have subsequently been fitted to a high-order polynomial or to a Vogel-Fulcher-Tammann form (the actual form of the function is not important as long as it fits well the data). From the HN or the GGE parametrizations, one can directly compute the associated relaxation function in the time domain~\cite{refBlochowicz2006and2003} and then obtain the three-point dynamic susceptibility $\chi_{T}(t)$ by simply taking the derivative of the fitted temperature parametrization.

In the case of the other experimental probes that we have considered, the data is directly provided in the time domain. For light and neutron scattering, we have fitted the relaxation function to the standard Kohlrausch stretched-exponential (KWW) function, whereas for the optical Kerr effect, we have used the parametrization given in Ref.~\cite{BPM}. As for the dielectric data, we have interpolated the temperature dependence of the parameters by polynomial or Vogel-Fulcher-Tammann functions and used the resulting formulas to compute the derivative with respect to temperature.

In our operational protocol, we strictly limit ourselves to the domains of measurement, avoiding extrapolations as well as combinations of several sets of data.

\subsection{Robustness of the temperature derivatives}
\label{denis}
Provided one does not extrapolate out of the experimentally accessed temperature domain, the above  parametrization scheme provides a very good description of the temperature and time dependences of the measured relaxation functions. However, one must recall that we are ultimately interested in a derivative with respect to temperature of a quantity whose characteristic timescale varies by many orders of magnitude for small temperature changes. One must therefore study the robustness of the outcome.

\begin{figure}
\psfig{file=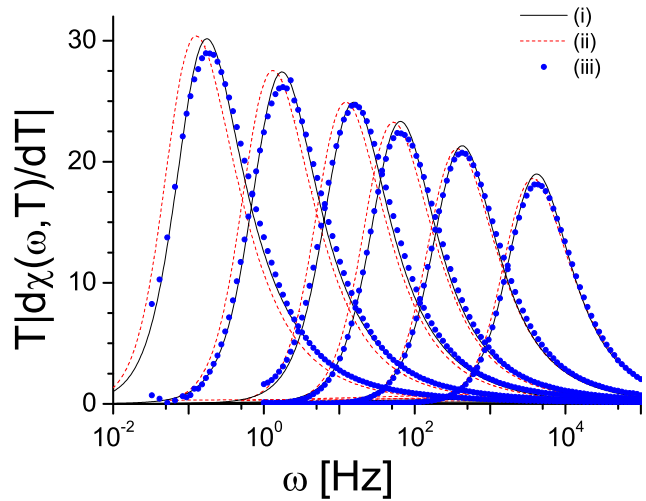,width=8.5cm}
\caption{\label{fig1}(Color online)
Three different estimates (i-iii) detailed in Sec.~\ref{denis}
for the full curve $T \partial \chi(\omega, T)/ \partial T$ for glycerol at temperatures
$T=196.6$, 201.6, 207.1, 211.1, 217.2, and 225.6~K
(from left to right). Full lines are for method (i), 
dashed lines for method (ii), and points for method
(iii). The three methods give consistent results, with small deviations that are well understood.}
\end{figure}

As a first check, we have focused on the dielectric relaxation of glycerol. We have measured the linear dielectric susceptibility of glycerol every 1~Kelvin for a set of temperatures above $T_{g} \approx 190\ $K
and below $232\ $K~\cite{refPise2006Ladieu}. From the measured complex capacitance,
 $ C( \omega ,T)$, we have obtained the following normalized dynamical
quantity:
\be
\chi(\omega,T) = \frac{\epsilon '( \omega ) - \epsilon _{\infty}}{\epsilon  ( 0 ) - \epsilon_{\infty}}  =
{\rm Re} \left[ \frac{C( \omega ) - C( \infty ) }{ C(0) - C( \infty )} \right],
\ee
where where $\epsilon ' ( \omega ) $ is the real part of the dielectric susceptibility. Contrary to $\epsilon (0)$,  $\epsilon _{\infty}$ has not been directly measured and has instead been obtained from a fit of
$C( \omega ,T)$ using the HN parametrization. It turns out that $\epsilon _{\infty}/ \epsilon (0)$ is a small number whose temperature dependence is sufficiently weak to be irrelevant in the calculation of the temperature derivative $\partial \chi(\omega,T)/\partial T$. We have computed the latter by three different methods. The two first ones are as described in Sec.~\ref{fitting}: $\chi(\omega)$ has been fitted to (i) the HN and (ii) the GGE forms, and the temperature dependence of the parameters has been fitted as well; $\partial \chi(\omega,T)/\partial T$ has then been computed from the parametrization of the $T$ and $\omega$ dependences. In the procedure (ii), we have directly taken the parameter values given by Blochowicz \textit{et al.}~\cite{refBlochowicz2006and2003} from adjustment to their own dielectric measurements on glycerol. The third method (iii) consists in a direct calculation of $\partial \chi(\omega,T)/\partial T$ through finite differences on the experimentally measured capacitance: $\partial \chi(\omega,T)/\partial T \simeq \left[ \chi(\omega,T+\frac{\Delta T}{2})-\chi(\omega,T-\frac{\Delta T}{2})\right] /\Delta T$, with $\Delta T$ a small, finite temperature step. Note that we have considered the frequency domain instead of the time domain because this allows a direct use of the raw data when computing the finite differences, without further manipulations.

The three different estimates obtained through (i-iii) are shown for several temperatures in Fig.~\ref{fig1}. The results are very similar. At a given $T$, $\partial \chi(\omega,T)/\partial T$ reaches a maximum, $\partial \chi/\partial T\vert^{\ast}$, at a frequency $\omega ^{\ast} $ which is close to the frequency $\omega _{\alpha} \sim \tau_\alpha^{-1}$ at which the imaginary part $ \epsilon '' ( \omega )$ of the dielectric susceptibility is maximum.
The values $\omega^\ast$ and $\partial \chi/\partial T\vert^{\ast}$ are very close for the three methods, with small differences which we now discuss.

The comparison of methods (i) and (iii) reveals that performing finite differences of raw data does not change the
value of $\omega ^{\ast}$, but slightly underestimates the maximum  $\partial \chi/\partial T\vert^{\ast}$ by an
amount which increases with $\Delta T$. For $\Delta T = 1\ $K, the maximum value obtained through (iii) is lower than the two others by about $4 \%$. The error on $\partial \chi(\omega,T)/\partial T$ due to our experimental uncertainties on the dielectric measurement is estimated to vary from less than $1 \%$ at the maximum $\partial \chi/\partial T\vert^{\ast}$ to $5 \%$ when $T \partial \chi(\omega,T)/\partial T$ is ten times smaller.
In Fig.~\ref{fig1}, it is worth noticing that method (iii) yields smooth curves despite the fact that no fitting procedure
of the data is involved. This is consistent with  errorbars of at most $1 \%$ at the frequency $\omega^\ast$.

As for the comparison between methods (i) and (ii) the difference between the
values of  $\partial \chi/\partial T\vert^{\ast}$ is small, less than  $3 \%$, as shown in Fig.~\ref{fig1}.
The main difference lies in the value of $\omega ^{\ast} (T)$. 
Our values (following (i)) are smaller than those of
Blochowicz \textit{et al.}~\cite{refBlochowicz2006and2003} by typically $20 \%$. This amounts to decreasing all our temperature values by $0.6\ $K. Beyond a mere problem of thermometer calibration, this discrepancy may come from a difference in the purity of the glycerol samples, in particular their water content. It has been shown~\cite{refRyabov2003} that a glycerol sample protected from any humidity yields, at a given $T$, values of $\tau _{\alpha}$ larger than those of a sample ``regularly exposed'' to air humidity, corresponding approximately to a temperature shift of $1\ $K, at least in the limited temperature interval considered here. This is quite larger than
the difference reported in Fig.~\ref{fig1}. To reinforce this interpretation, let us mention that in another series of experiments (not reported here) we have found values of $\omega ^{\ast} (T)$ closer to those of Ref.~\cite{refBlochowicz2006and2003}. The difference between our two sets of data amounts to a $0.17\ $K shift, while the exact same thermometer was used in both experiments. The important point for our present purpose is that the difference in $\omega ^{\ast} (T)$ does not come with any significant difference in the maximum value of  $\partial \chi/\partial T\vert^{\ast}$.

These detailed experiments performed on glycerol show that the temperature derivative involved in
$\Ncorr$ can be consistently obtained by using either finite differences (as also done in numerical work~\cite{I}),
or a parametrization of the temperature and time (or frequency) dependences of the relaxation data through the
HN or GGE descriptions.

\subsection{Dependence of $N_{{\rm corr},T}$ on thermodynamic inputs}

\begin{figure}
\psfig{file=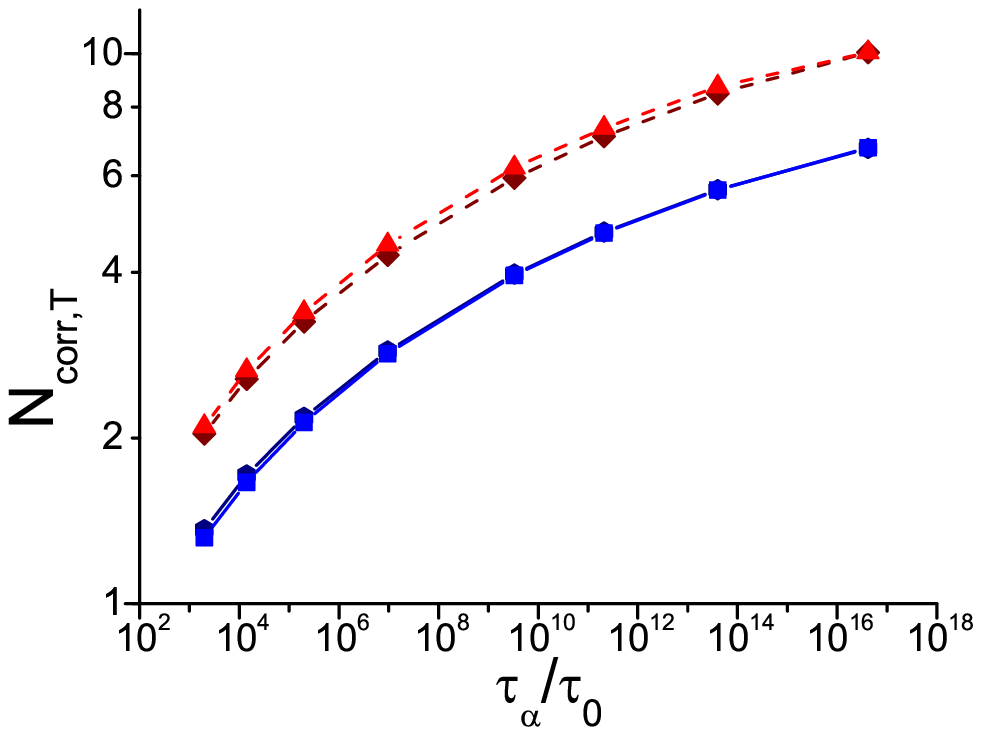,width=8.cm}
\psfig{file=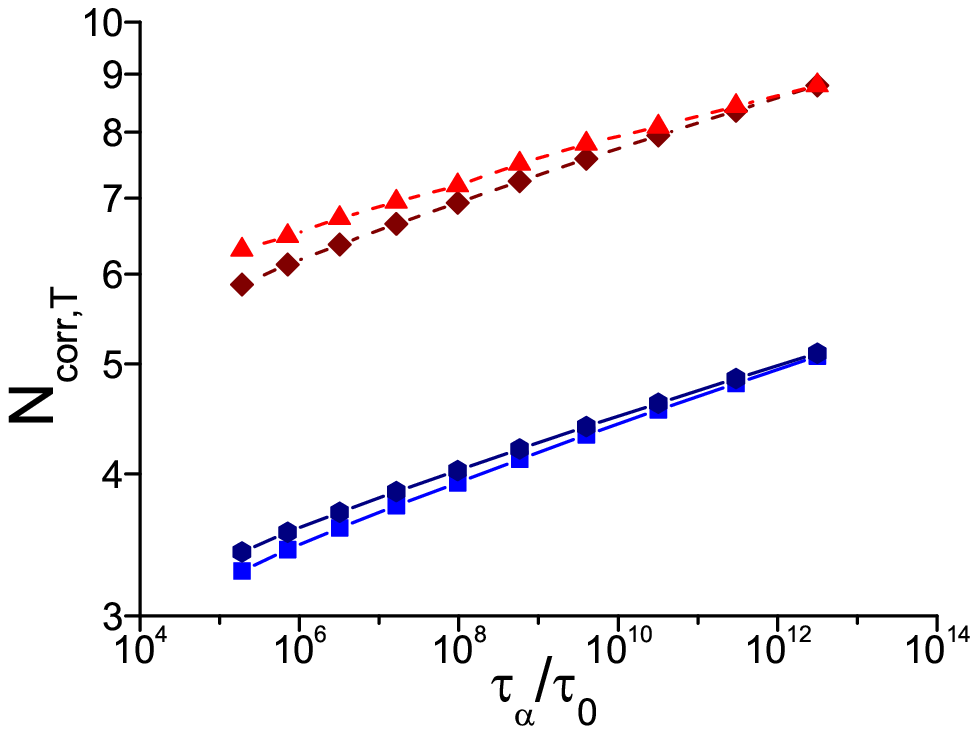,width=8.cm}
\caption{\label{fig2} (Color online)
$N_{{\rm corr},T}$ versus $\tau_\alpha/\tau_0$ on a log-log plot for 
glycerol (top) and \textit{o}-terphenyl (bottom).  
$N_{{\rm corr},T}$ is calculated by using $c_p(T_g)$ (circles), $c_p(T)$ (squares), $\Delta c_p(T_g)$ (diamonds) and $\Delta c_p(T)$ (triangles). The two upper (resp., lower) curves correspond to using $\Delta c_p$ (resp., $c_p$);
$\tau_0$ is arbitrarily set to 1~ps.}
\end{figure}

As discussed in Sec.~\ref{assessment:section} and illustrated by 
Eq.~(\ref{eq10:equation}), determining $N_{{\rm corr},T}$
(and consequently $N_{{\rm corr},4}$) requires thermodynamic input, primarily the isobaric specific heat $c_{p}$.
For a few representative glassforming liquids for which enough data are available, we have investigated the two
following points: (1) the quantitative difference in the estimate of $N_{{\rm corr},T}$ and its temperature
dependence which is introduced by using either $c_{p}$ or $\Delta c_{p}$, the value in excess to that of the
associated crystal (see the discussion in Sec.~\ref{from}); (2) the effect on the $T$-dependence of  $N_{{\rm corr},T}$
that results from replacing  $c_{p}(T)$ (or $\Delta c_{p}(T)$) by a constant value, taken at $T_{g}$, $c_{p}(T_{g})$
(or $\Delta c_{p}(T_{g})$). There are indeed a number of glassforming materials for which only the latter value is known.

The results of our comparative study on two systems, glycerol~\cite{Cpgly} and \textit{o}-terphenyl~\cite{CpoTP},
are displayed in Fig.~\ref{fig2}, where we plot $N_{{\rm corr},T}$ as a function of $\tau_{\alpha}/ \tau_{0}$,
where $\tau_{\alpha}$ is the $\alpha$-relaxation time extracted from the HN fit of the dielectric data and $\tau_{0}$ is
an arbitrarily chosen microscopic time of $1$ ps. One observes that the evolution is essentially the same in the different
cases, using either $c_{p}(T)$, $\Delta c_{p}(T)$, $c_{p}(T_{g})$, or $\Delta c_{p}(T_{g})$. Replacing $\Delta c_{p}(T)$ by a constant $\Delta c_{p}(T_{g})$ leads to a maximum deviation in $N_{{\rm corr},T}(T)$ of about $10-15\%$ at high temperature around the melting point. Obviously, using $\Delta c_{p}$ in place of $c_{p}$ increases the absolute value of $N_{{\rm corr},T}$, by up to a factor 2 in the case of \textit{o}-terphenyl, but the whole shape of $N_{{\rm corr},T}(T)$ is not significantly altered. Therefore, in what follows, we shall 
present results calculated with $\Delta c_{p}(T_{g})$, or, when
not available, with $c_p(T_g)$.

\subsection{Dependence of $N_{{\rm corr},T}$ on the experimental probe}

\label{probe}

In our estimates of $\Ncorr(T)$ we have focused on dielectric spectroscopy for which the largest
data base is available. One may however inquire how much would the results change if other experimental
probes were used. On the one hand, one expects that spatial correlations of the dynamics are an intrinsic
property of the material, irrespective of the probe monitoring the dynamics. On the other hand, the
estimates provided by Eq.~(\ref{eq10:equation}) and (\ref{eq11:equation}) are not as straightforward
as one would like (see the discussion in Secs.~\ref{from}, \ref{jp-discussion}) and they may depend on the probe chosen. This can
be simply illustrated in the case where the relaxation functions obey ``time-temperature superposition''.
As shown by Eq.~(\ref{eq12:equation}), $N_{{\rm corr},T}$ is proportional to $\vert \Phi'(1)\vert$, 
which is essentially
the stretching parameter $\beta$ (it is equal to $\beta/e$ if $\Phi(x)$ is a stretched exponential,
 $e^{-x^\beta}$). If two probes are characterized by the same $T$-dependence of their relaxation time but by
  different stretching parameters, which is not uncommon (see \textit{e.g.} Ref.~\cite{brodin}), the resulting
  estimates of $N_{{\rm corr},T}$ then differ: the more stretched the relaxation (smaller $\beta$), the smaller the value of $N_{{\rm corr},T}$.

\begin{figure}
\begin{center}
\psfig{file=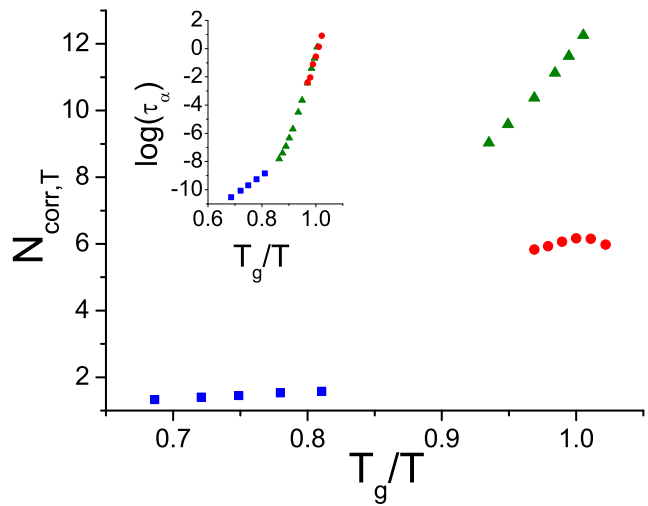,width=8.5cm}
\caption{(Color online) 
$N_{{\rm corr},T}$ versus $T_g/T$ for \textit{m}-toluidine evaluated from dielectric (triangles) and photon correlation (circles) spectroscopies close to $T_g$, and by coherent quasi-elastic neutron scattering (squares) at high temperature. (We have not considered the two lowest temperature data points
for which the determination of the relaxation profile is too uncertain.)
The inset shows the corresponding relaxation times (determined from the 
maximum of the imaginary part of the dielectric susceptibility, from 
a KWW fit of the correlation functions obtained from 
photon correlation spectroscopy and neutron scattering at $q=1.3$~\AA$^{-1}$).
\label{fig4}}
\end{center}
\end{figure}

Matter can be worse if time-temperature superposition does not hold.
We illustrate this point by considering several sets of relaxation
data for liquid $\textit{m}$-toluidine: low-temperature (near
$T_{g}$) dielectric~\cite{OTP} and photon correlation spectroscopy
(PCS)~\cite{aouadi} measurements, high-temperature (above and around
melting) neutron scattering data. Over the limited
domains of temperature studied, both dielectric~\cite{OTP} and neutron
scattering relaxation functions appear to satisfy time-temperature
superposition with (accidentally) almost the same stretching
exponent $\beta \simeq 0.6$ (the dielectric Kohlrausch stretching exponent is estimated from 
the fit in frequency domain according to ~\cite{lindsey}). On the other hand, the PCS data show a
marked increase of the stretching, \textit{i.e.} a decrease of
$\beta$, as $T_{g}$ is approached~\cite{aouadi}. The resulting
estimates of $N_{{\rm corr},T}$ are shown in Fig.~\ref{fig4} . 

The PCS data lead to a smaller value of $N_{{\rm corr},T}$ than the dielectric ones (by a factor of about 2) and 
to a slower growth as one lowers the temperature near $T_g$. 
This behavior comes from the competition between the 
increase of $\log \tau_{\alpha}$ and the decrease of $\beta$ as $T$ decreases. Similar results could be expected 
for other molecular liquids due to analogous differences between dielectric and light-scattering dispersions 
\cite{OTP,puzzling}.

The observation that average dynamic relaxations might 
depend on the probe is one of the puzzling
experimental features of the glass transition. One usually
gets away by arguing that $\tau_\alpha$ from different
probes are nevertheless consistent with one another, while different shapes
result from different microscopic observables (when they can be 
properly defined). This becomes even more puzzling when     
these data are derived to obtain the value of $N_{{\rm corr},T}$. 
If the observed averaged behaviors 
do not coincide, then it is no surprise that 
derivatives look even more different, as seen in Fig.~\ref{fig4}. 
But this raises interesting questions if one wants to interpret
the result in terms of a correlation volume: 
are there experimental probes that are ``better'' than other 
because more directly coupled to molecular degrees of freedom? 
Can the high precision obtained in the dielectric 
spectroscopy experiments of Sec.~\ref{denis} be obtained 
using other probes? Is the discrepancy 
observed in Fig.~\ref{fig4} due to the comparison 
of data with different resolutions?
Can different microscopic observables 
be correlated over different lengthscales? 
Is it possible to extract a probe-independent correlation volume?
More work is certainly needed to answer these questions.

\section{Temperature dependence of $\Ncorr$ for a range of glassformers}

\label{alldata}

\begin{figure}
\psfig{file=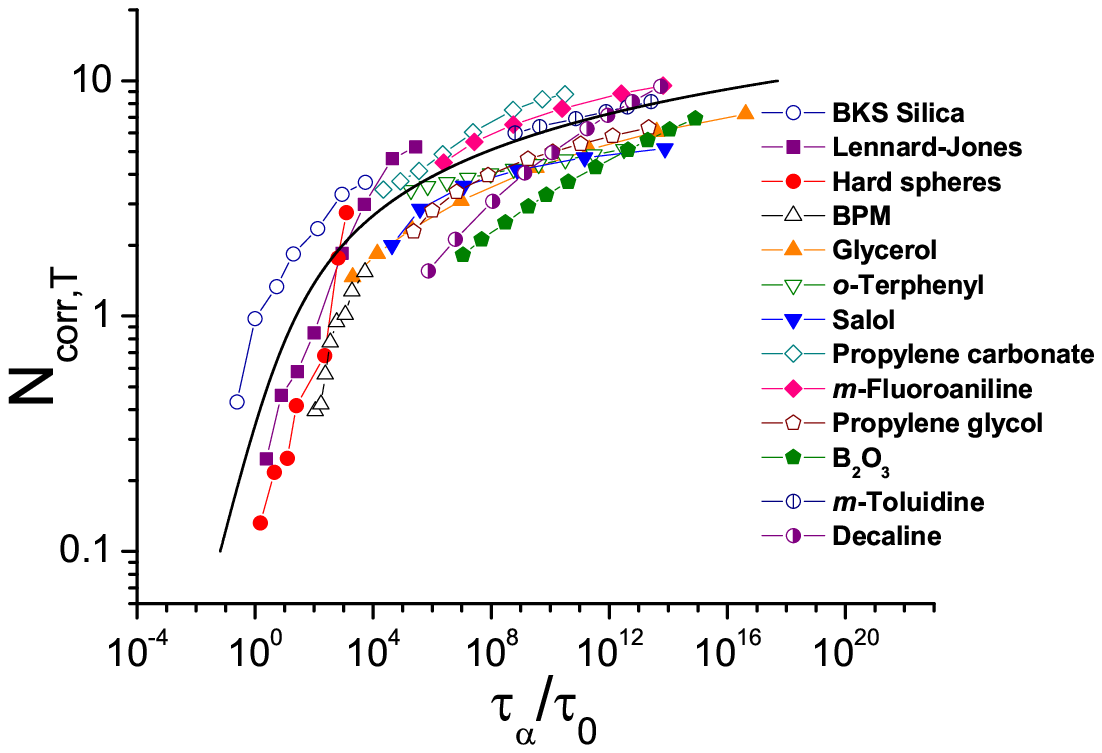,width=8.5cm}
\psfig{file=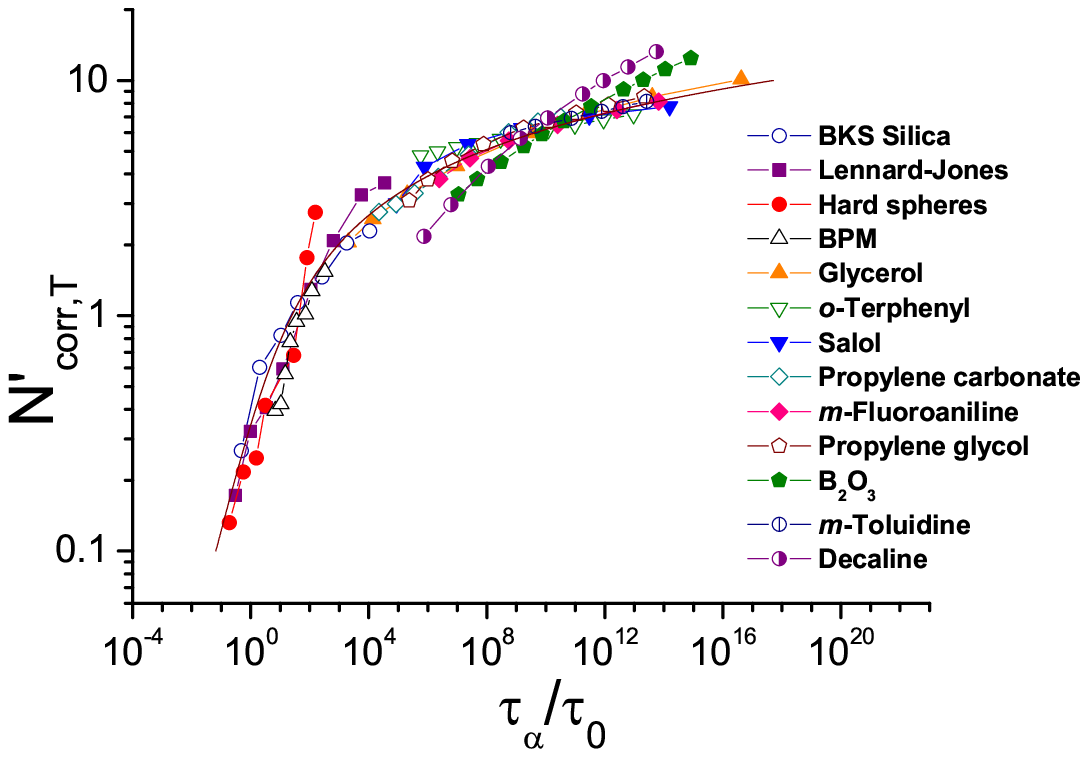,width=8.5cm}
\caption{(Color online)
Top: Evolution of $N_{{\rm corr},T}$ for the different
materials indicated when the glass transition is approached. The
results are shown on a logarithmic scale as a function of
$\tau_{\alpha}/ \tau_{0}$, where $\tau_{0}$ is a microscopic
timescale (see the text). The full line is from Eq.~(\ref{eq:fit})
and describes a crossover from a power law scaling at high
temperatures (small $\tau_\alpha$) to a logarithmic growth close to the glass transition, Eq.~(\ref{eq:fit}). Bottom: Using the freedom left by unknown
normalizations of order unity we obtain a better 
collapse of the data onto the fit.
\label{fig5}}
\end{figure}

\begin{figure}
\psfig{file=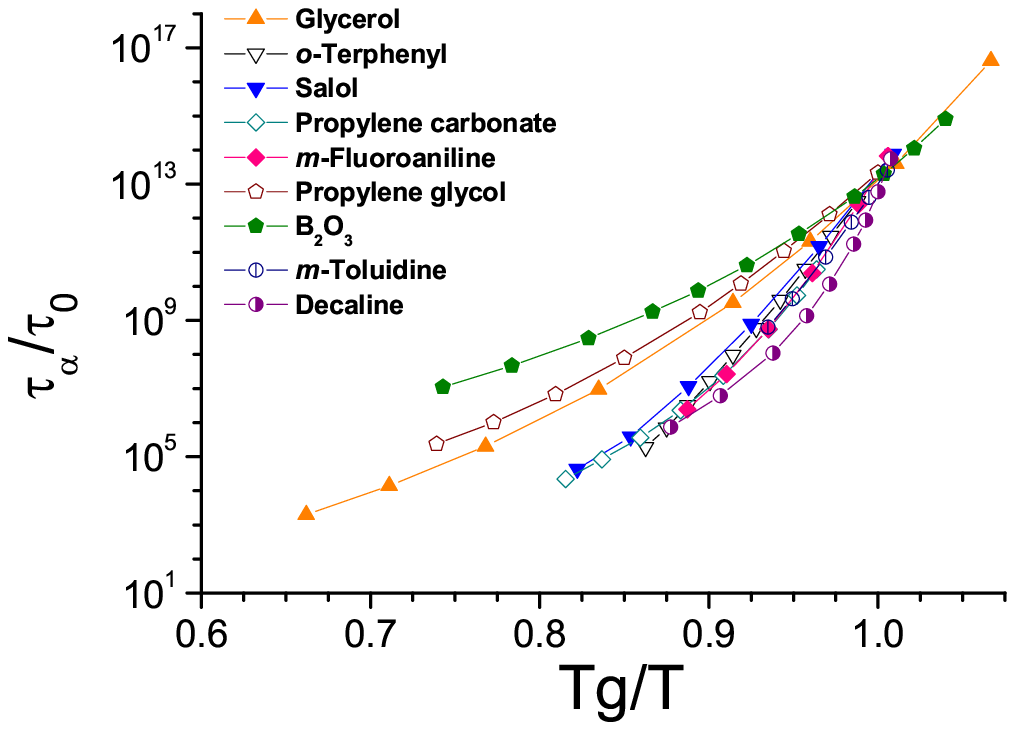,width=8.5cm}
\psfig{file=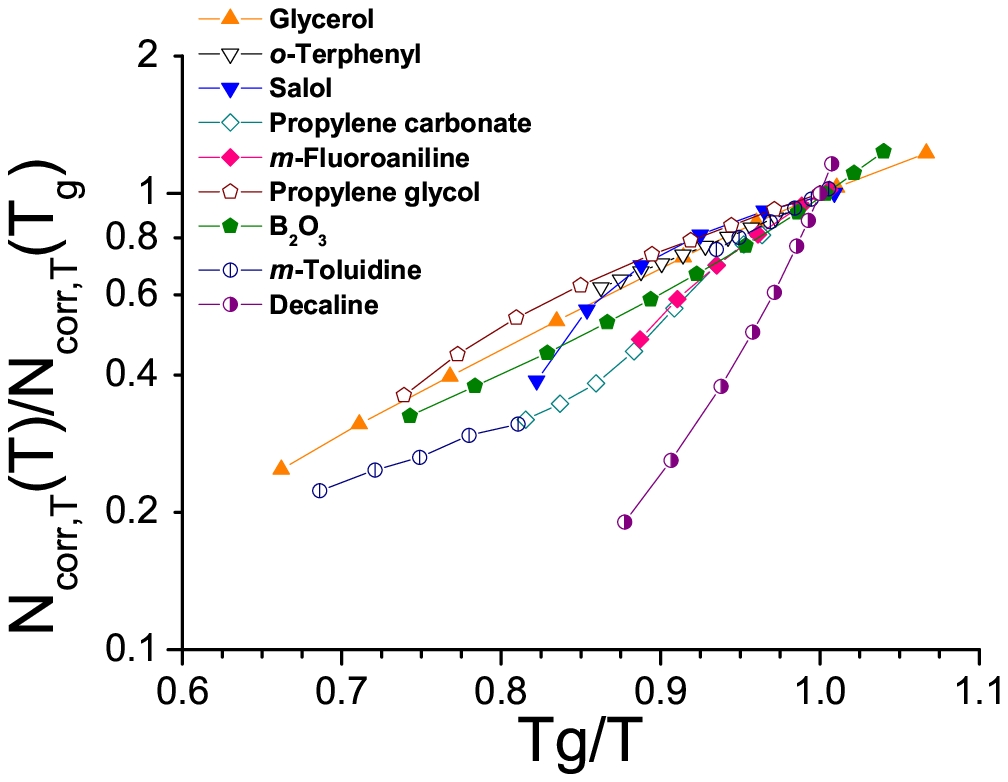,width=8.5cm}
\caption{(Color online)
$N_{{\rm corr},T}/N_{{\rm corr},T}(T_g)$ (top) and $\tau_{\alpha}/\tau_0$ (bottom)   
versus $T_g/T$ for different glass-forming liquids (indicated in the Figure) 
on a logarithmic scale;
$T_g$ is defined by $\tau_\alpha(T_g)=10$~s.
\label{fig6}}
\end{figure}

Having detailed how to estimate from experimental data the
temperature dependence of the number of molecules whose dynamics are
spatially correlated and discussed the robustness of
our method, we now present the results for a variety of 
glassforming liquids.
The data shown in
Figs.~\ref{fig5} and \ref{fig6} 
represent the central outcome of our work.

In Fig.~\ref{fig5} (top) we plot on a logarithmic
scale $N_{{\rm corr},T}$ determined using Eq.~(\ref{eq10:equation}) 
versus $\tau_{\alpha}/ \tau_{0}$,
where $\tau_{\alpha}$ is the $\alpha$-relaxation time and $\tau_{0}$
is a microscopic time arbitrarily fixed to $1$ ps (except for a
couple of systems discussed below). 
Due to the absence of data for $\Delta c_p$ for the last three systems,
results for all substances in Fig.~\ref{fig5} are computed using $c_p(T_g)$.
The data include glycerol,
\textit{o}-terphenyl~\cite{OTP}, $m$-toluidine~\cite{OTP}, salol~\cite{salol}, propylene carbonate~\cite{refBlochowicz2006and2003}, \textit{m}-fluoroaniline~\cite{refBlochowicz2006and2003}, propylene glycol~\cite{refBlochowicz2006and2003}, and decaline~\cite{decaline} (dielectric measurements), as well as B$_2$O$_3$~\cite{B2O3} (photon correlation spectroscopy), BPM~\cite{BPM} and salol~\cite{salolOKE} (optical Kerr effect), and $m$-toluidine (neutron scattering). We have added for comparison simulation results obtained on two very different model glassformers, namely the BKS model for silica~\cite{I} and a binary Lennard-Jones mixture~\cite{I}. In those two cases, the correlation functions used to probe the dynamics are self-intermediate scattering functions, and temperature derivatives are measured using finite differences of measurements performed at nearby temperatures, $T$ and $T+\delta T$, with $\delta T$ small enough that linear response applies. For BKS silica, we take $\tau_0=1$~ps, while we take the standard Lennard-Jones time unit as a microscopic timescale, $\tau_0= \sqrt{m \sigma^2/(48\epsilon)}$, where $m$, $\sigma$ and $\epsilon$ respectively represent the mass of the particles, their diameter, and the depth of the Lennard-Jones potential; using argon units one gets $\tau_0 \approx 0.3$~ps. Finally, we have included a colloidal hard-sphere system, simply reporting the results already published in Ref.~\cite{science} and choosing $\tau_0=1$~ms.

A key feature in the results presented in Fig.~\ref{fig5} 
is that $N_{{\rm corr},T}$ (and via Eq.~(\ref{eq11:equation}), $N_{{\rm corr},4}(T)$) does increase as temperature decreases or the relaxation 
timescale increases. 
The extent of spatial correlation in the dynamics therefore grows as one
approaches the glass transition~\cite{science}. 

Although there is an unfortunate shortage of data covering the whole
temperature range from above melting down to $T_{g}$,
Fig.~\ref{fig5} clearly indicates that $N_{{\rm corr},T}$
grows faster when $\tau_{\alpha}$ is not very large. The
initial, fast growth observed in $\Ncorr$ occurs at high
temperature, close to the onset of slow dynamics. A power law fit, $N_{{\rm corr},T} \sim \tau_\alpha^{1/\gamma}$,
in this regime ($\tau_\alpha /\tau_0 < 10^6$) leads to values $\gamma=2-3$. Interestingly, these 
are consistent with the predictions of Mode Coupling Theory, 
see \cite{II,BBMR}. 
The growth becomes much slower as $T_{g}$ is approached. More quantitatively, if one
wants to fit with a (second) power law 
in the regime $\tau_\alpha /\tau_0 \in [10^6, 10^{12}]$, one finds
$N_{{\rm corr},T} \sim \tau_\alpha^{1/\gamma'}$, with $\gamma' > 20$ 
and small variations
among the different liquids. This means that when the dynamics slows
down by 6 decades in time, $N_{{\rm corr},T}$ merely increases by a
factor about 2. This slow growth is actually broadly
consistent with activation-based theoretical approaches which all
predict some sort of logarithmically slow growth of dynamic
lengthscales~\cite{rfot,Gilles,arrow,droplet}. In order to capture the idea
of a fast initial growth of $N_{{\rm corr},T}$ when $ \tau_\alpha$
is small, followed by a logarithmically slow growth, we have
empirically fitted the data to the formula:
\begin{equation}
\tau_\alpha \simeq A \left( \frac{N_{{\rm corr},T}}{N_0} \right)^\gamma \exp \left[
\left( \frac{{N_{{\rm corr},T}}}{N_0}\right)^\psi \right],
\label{eq:fit}
\end{equation}
and found rough agreement (see the full line in Fig.~\ref{fig5}) with $A=4$, $N_0 = 0.8$, $\gamma=2$,
and $\psi=1.4$. (Note that very similar ideas have been used in spin glass studies~\cite{SG}.)
These values are unfortunately only indicative because a set of rather different values allows one to fit 
almost equally well the curves. It would be interesting to be able to narrow down the range of acceptable values
for $\psi$ in order to test more stringently the various activation-based theories found in the literature~\cite{rfot,Gilles,arrow,droplet}.

A second interesting feature of the results presented in Fig.~\ref{fig5}
is that the data for different materials 
fall remarkably close to one another, despite 
the various caveats mentioned above about prefactors, normalization etc., which
might affect the vertical axis in this figure. 
A signature of such caveats is the fact that we find 
values of $N_{{\rm corr},T}$
at high temperature which fall below 1. 
Investigating in detail the 
reasons of this
behavior and improving the normalization procedure, 
however, would require the
knowledge of the spatial distribution of the 
dynamical correlations, which is for the moment out of reach experimentally.
Moreover our choice  
of a microscopic timescale (mainly 1~ps) is also somewhat arbitrary. 
Clearly, a better collapse of the different curves could be obtained 
by making use of the freedom offered by these unknown normalizations. 
In Fig.~\ref{fig5} (bottom) we rescale vertical and horizontal axis 
by factors or order unity ({\it i.e.} we only 
allow data shifts within a decade, very often much less) in order to obtain 
a better collapse along the fit described above. We find that
all data collapse rather well, suggesting that
Eq.~(\ref{eq:fit}) captures the physics of glassformers very well.

We show in Fig.~\ref{fig6} (top) the same data as in Fig.~\ref{fig5},
normalized by the value of $N_{{\rm corr},T}$ at $T_{g}$, versus
$T_{g}/T$ (excluding the simulation results and the colloidal
system, for which $T_{g}$ cannot be determined).
Table~\ref{tab1} shows the different values of $N_{{\rm
corr},T}(T_{g})$. Recall that estimates for $N_{{\rm corr},4}(T)$
are simply obtained through Eq.~(\ref{eq11:equation}), which on a
logarithmic scale merely amounts to rescaling the $y$-axis by a
factor of $2$.

\begin{table}
\begin{tabular}{|c|c|c|c|c|}
\hline
glassformer&$N_{{\rm corr},T}(T_g)$&$N_{{\rm corr},4}(T_g)$ & 
$\Delta c_p$ & $c_p$ \\
\hline
glycerol & 8.2 & 67.2 & 90.5 & 175 \\
\textit{o}-terphenyl & 9.2  & 84.6 & 112.7 & 333.7 \\
salol & 8.6 & 74.0 & 118 & 320 \\
propylene carbonate & 15.9 & 252.8 & 75.4 & 164 \\
\textit{m}-fluoroaniline & 12.9 & 166.4 & 86 & 161 \\
propylene glycol & 9.4 & 88.4 & 67.2 & 150 \\
$B_2O_3$ & 10.1 & 102.0 & 40 & 131 \\
\textit{m}-toluidine & 11.7  & 136.9 & 89.5 & 192.6 \\
decaline & 11.8  & 139.2 & 64 & 133 \\
\hline
\end{tabular}
\caption{Values of $N_{{\rm corr},T}(T_g)$ for various glassforming liquids, 
and corresponding lower bound on $N_{{\rm corr},4}(T_g)$.  
The values are computed with $\Delta c_p(T_g)$. For completeness
we report the values of $c_p$ and $\Delta c_p$ at $T_g$ in 
J.~K$^{-1}$.~mol$^{-1}$.
\label{tab1}}
\end{table}

In Fig.~\ref{fig6} we also compare the celebrated
Angell plot (bottom) that illustrates the relative fragility of
glassformers~\cite{angell} to the relative change of $N_{{\rm corr},T}(T)$ (top). We do not find any systematic correlation between the two plots,\textit{ i.e.}, between the rapidity of the increase of $N_{{\rm corr},T}(T)$ and the fragility (with the notable exception of decaline which is also the most
fragile material considered). Indeed, the results for all the systems show a rather similar behavior, characterized above all by a fairly modest growth of $N_{{\rm corr},T}$.
We note in passing that since $\Delta c_p$ is 
taken here as constant and replaced by its value
at $T_g$, Eq.~(\ref{eq12:equation}) 
would on the contrary predict that $\Ncorr(T)$ 
follows the fragility pattern.
The fact that this is not what is observed 
is mainly due to the breakdown of the time-temperature
superposition property and the resulting effect of the 
shape (stretching) factor.

Finally, and keeping in mind the theoretical and practical uncertainties on the determination of the absolute
 value of $\Ncorr$, we note that the values of $N_{{\rm corr},T}(T_{g})$ for all studied glassformers are similar,
all between 8.2 and 15.9 as shown in Table \ref{tab1}. Contrary to what has been done in Ref.~\cite{science},
we have not attempted here to convert this number into a precise length scale (see the discussion above), nor to 
``normalize'' $N_{{\rm corr},T}(T_{g})$ (or $N_{{\rm corr},4}(T_{g})\simeq N_{{\rm corr},T}(T_{g})^{2}$)
by using an effective number of ``beads'' per molecule to account for the difference in complexity of the various molecules.
The relative 
independence of $\Ncorr(T_g)$ upon fragility is broadly consistent with 
Wolynes's version of the Random First Order Theory~\cite{rfot,rfot2}, 
although the
relation between $\Ncorr$ and the typical size of the mosaic state is 
far from obvious -- see the conclusion for a deeper discussion of this point.

\section{Relative effects of temperature and density fluctuations on dynamical correlations}

The last question we want to address is the relative effects of temperature 
and density fluctuations on dynamical correlations. In Sec.~\ref{2C}, 
we have shown that the relative influence of temperature and
density fluctuations on  $\chi_{4}$ is represented respectively by the contributions $k_{B}T^{2}/c_{v}(\chi_{T}^{NVT})^{2}$ and $\rho^{3}k_{B}T\kappa_{T}(\chi_{\rho}^{NPT})^{2}$.
We have also argued in Sec.~\ref{jp-discussion} that both terms in fact represent a (squared) dynamic response and should behave similarly as a function of temperature, provided both $T$ and $\rho$
locally couple to the dynamics. It is therefore of interest to compare these two contributions, which we now do
for three molecular liquids 
characterized by different fragilities: glycerol, $m$-toluidine, 
$o$-terphenyl. 

\begin{figure}
\psfig{file=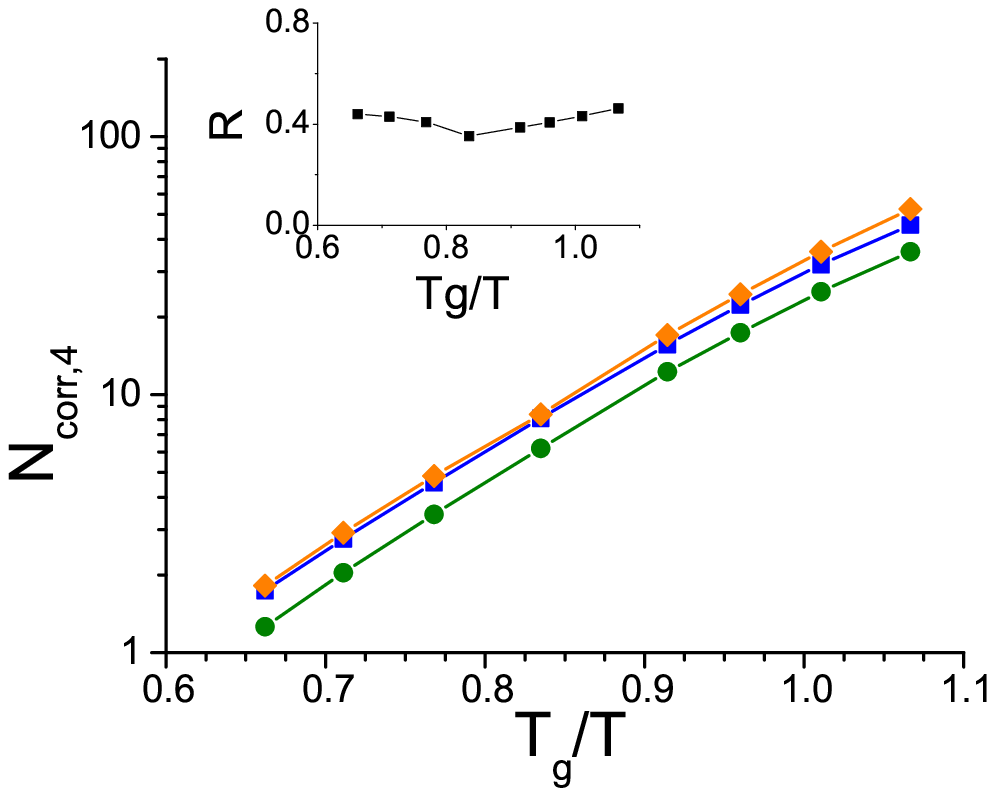,width=8cm}
\psfig{file=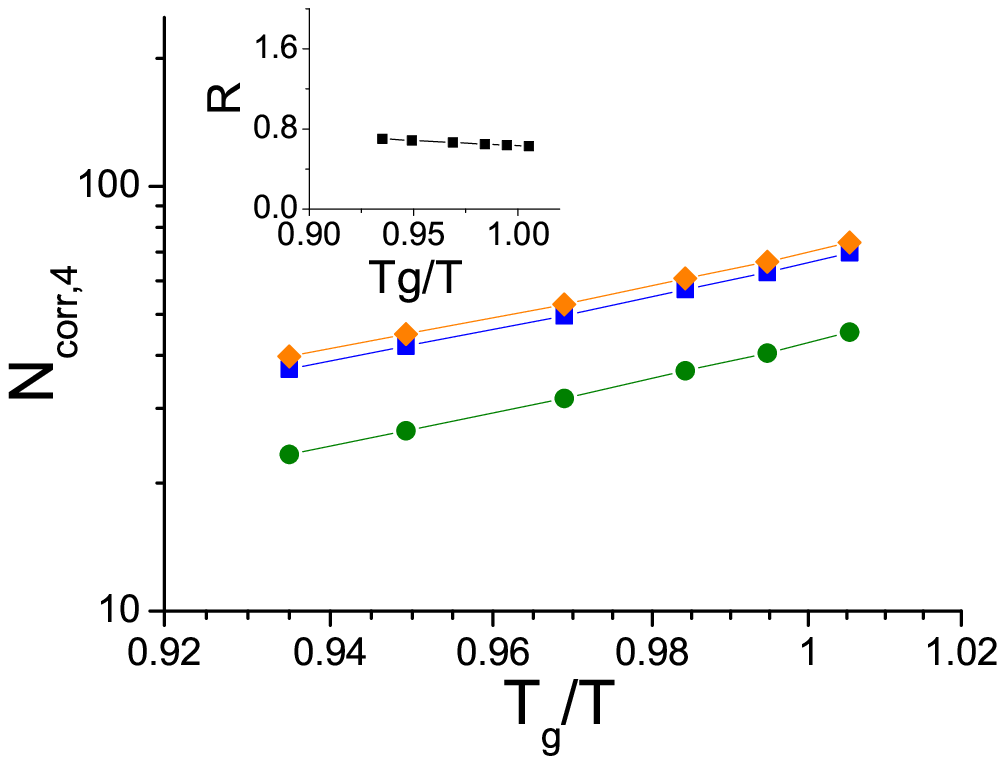,width=8cm}
\psfig{file=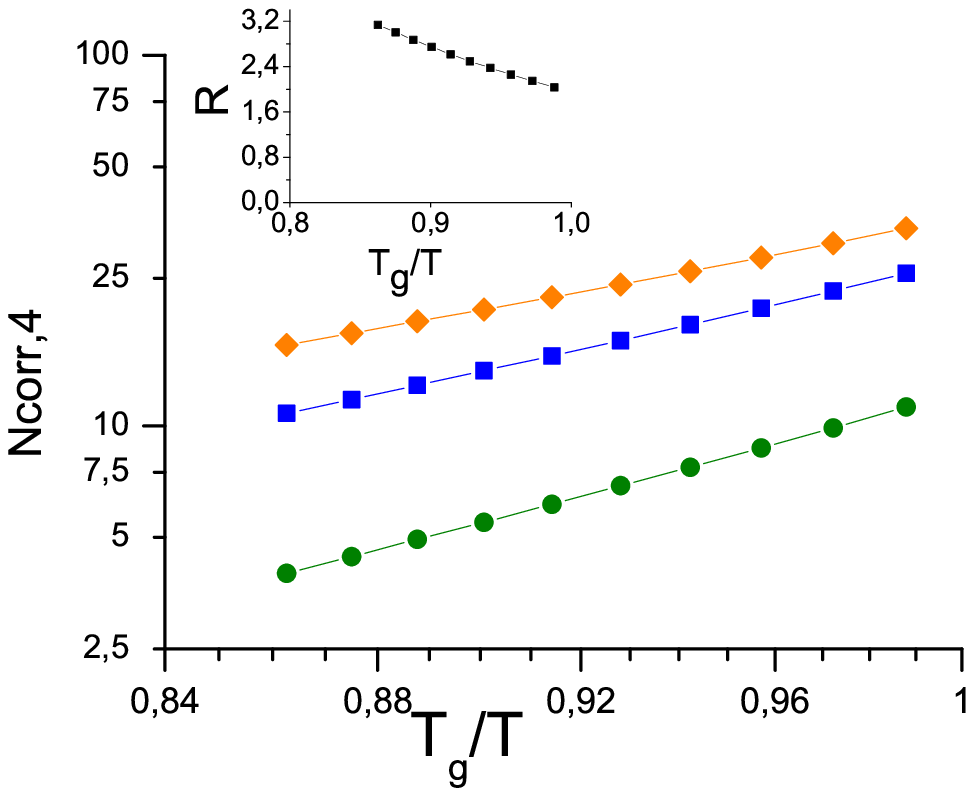,width=8cm}
\caption{(Color online)
$N_{{\rm corr},4}$ versus $T_g/T$ for glycerol, $m$-toluidine, and $o$-terphenyl (top to bottom) calculated via the lower bounds given in eq.~\ref{eq13:equation} (squares) and ~\ref{eq14:equation} (diamonds). Also displayed is the contribution due to energy fluctuations (circles). The inset shows the ratio R versus $T_g/T$ (see eq.~\ref{eq:equation21})\label{fig7} 
}
\end{figure}

We introduce a number of simplifications in our protocol. (1) We assume that the property of ``time-temperature superposition''
holds for the temperature range under study. This allows us to write the relaxation function as $F(t,T)=\Phi(t/\tau_{\alpha}(T))$
with $\Phi(0)=1$ and no explicit $T$-dependence elsewhere than in the relaxation time $\tau_{\alpha}(T)$. (2) We consider
the thermodynamic coefficients $c_{p}$, $c_{v}$, $\kappa_{T}$, not the excess with respect to the crystal values, which
usually are not all accessible. The data for $c_{p}(T)$ and the equation of state are taken from the
literature~\cite{CpoTP, Cpmtol, Cpgly, PVToTP, PVTmtol, PVTgly} and $c_{v}$ is obtained from the relation
$c_{v}=c_{p}-\frac{T\alpha_{p}^{2}}{\rho \kappa_{T}}$, where $\alpha_{p}$ is the isobaric coefficient of expansion.
(3) We use the scaling form onto which the density and the temperature dependences of the $\alpha$-relaxation time and
the viscosity of many glassforming liquids and polymers have been found to collapse~\cite{alba,roland},
\begin{equation}
\tau_{\alpha}(\rho,T) =  f[e(\rho)/k_{B}T],
\end{equation}
where $f(x)$ is a species dependent, scaling function and $e(\rho)$ is a density-dependent (activation) energy scale.
With the above simplifications, and dropping the irreducible contribution coming from $\chi_{4}^{NVE}$, one finds from Eq.~(\ref{eq14:equation}):
\begin{equation}
N_{{\rm corr},4}(\rho,T) \simeq  \Phi'(1)^{2}\dfrac{k_{B}m_{v}^{2}}{c_{v}}\left( 1+\rho T\kappa_{T}c_{v} x^{2}\right) ,
\end{equation}
where $m_v=-\partial \log\tau_\alpha / \partial \log T \vert_V$ is the
``isochoric fragility'', which depends on $\rho$ and $T$ only through
$\tau_{\alpha}$~\cite{alba} and $x(\rho)=\partial \log
e(\rho)/\partial \log \rho$ is found roughly constant over the range
of densities involved at atmospheric pressure~\cite{alba,roland}.
The ratio of the contribution of the density fluctuations over that
of the energy fluctuations is then given by:
\begin{equation}
R = \rho T\kappa_{T}c_{v} x^{2}.
\label{eq:equation21}
\end{equation}
As shown in the insets of Fig.~\ref{fig7}, we 
find that $R$ is roughly constant and close to 0.4 for glycerol,
varies as $T$ decreases from 0.7 to 0.6 for $\textit{m}$-toluidine
(note that in this case only values close to $T_g$ are reported),
and from 3 to 2 for \textit{o}-terphenyl.
These results are compatible with 
our expectation, detailed in Sec.~\ref{jp-discussion}, that density and
temperature contributions mirror the same physical phenomenon, namely the anomalous susceptibility of the 
dynamics to a local perturbation, and hence should share a 
similar temperature dependence.
Thus, although the temperature contribution is dominant the other 
contribution behaves similarly as the temperature is lowered. In particular the ratio of the two terms 
changes modestly  compared to their overall variation, 
except perhaps for  \textit{o}-terphenyl.
It is known that in the latter case, density plays a more important role, 
as noticed for the slowing down of the relaxation~\cite{roland}.
It is however much less so
in the other liquids where the contribution of the energy
fluctuations dominate, especially at low temperature.

As a byproduct of the present analysis, we can also evaluate the relative quality of the two lower bounds
of $\chi_{4}^{NPT}$  that are provided either by Eq.~(\ref{eq13:equation}) 
with $\chi_{4}^{NPH}$ neglected
or Eq.~(\ref{eq14:equation}) with $\chi_{4}^{NVE}$ neglected. The result for the three liquids already
considered above is shown in Fig.~\ref{fig7}. Because it takes
more properly into account the relevant sources of dynamic fluctuations, the lower bound calculated
with Eq.~(\ref{eq14:equation}) is better than that 
obtained from Eq.~(\ref{eq13:equation}). Although we
 do not have access to $\chi_{4}^{NPH}$ and $\chi_{4}^{NVE}$ individually, we can estimate their
 difference (evaluated as before for the time 
$t \sim \tau_{\alpha}$ at which $\chi_{4}^{NPT}$ is maximum).
 At $T_{g}$, we find that 
$(\chi_{4}^{NPH} - \chi_{4}^{NVE} )/(\chi_{4}^{NPT} 
- \chi_{4}^{NPH})$ is 0.15, 0.06, 0.31
 for glycerol,  $\textit{m}$-toluidine, and \textit{o}-terphenyl, respectively.

\section{Concluding remarks}
\label{conclusion}

Our central result has been to provide direct evidence from experimental data for an increase of spatial correlations in the dynamics of glassforming liquids as one lowers the temperature~\cite{science}. Although the precise relation to a lengthscale is not straightforward (as discussed in Sec.~\ref{from}), this nonetheless shows the existence of a length that grows as the glass transition is approached. Moreover, our results establish an important characteristic of this phenomenon: despite the dramatic slowing down of the relaxation in supercooled liquids, the growth of $\Ncorr$, and of the associated lengthscale, is fairly modest, especially below the onset of slow dynamics. This is compatible with an activation-based picture of the dynamics in supercooled liquids. At higher temperature, however, the growth of this lengthscale appears to
be faster, and not inconsistent with the predictions of Mode-Coupling Theory. 

The modest size reached by the correlation volume at $T_g$ also conveys a two-fold message for future studies on the glass transition. 
(1) Experiments will presumably never be able to probe lengthscales as large as the ones measured for standard critical phenomena (where time and 
length scales are generically related in a power law fashion \cite{HalperinHohenberg}). As a consequence, it is unlikely that 
the growing of the lengthscale may provide {\it indisputable} evidence that the slowing down is driven by an underlying phase transition, whatever its nature.
Similarly, theoretical results that neglect subleading contributions to the asymptotic critical behavior will have to be taken with a grain of salt. (2) Despite the huge gap in timescales between experimental and numerical work, the probed lengthscales are not very different, which indicates the ability of numerical simulations to capture some major aspects of the slow dynamics in glassforming materials. A similar point was made in the context of 
spin-glasses~\cite{SG,SG2}. 

We have already stressed in this article the direct connection between  $\Ncorr$ and the dynamic heterogeneities observed in glassforming systems: $\Ncorr$ can be taken as the number of molecules participating in a typical dynamic heterogeneity. One may however wonder if $\Ncorr$ and the associated lengthscale relate to other aspects of the slowing down of relaxation leading to the glass transition. Among these aspects are usually mentioned the ``cooperative'' nature of the $\alpha$-relaxation~\cite{DS,walter,AG} and the stretching of the relaxation functions, often associated with a spread of local relaxation times~\cite{ediger,sillescu,richert}.

A connection between $\Ncorr$ and what is usually implied by ``cooperativity'' seems to us dubious, and certainly more work is needed to clarify 
this issue. 
Cooperativity, in the context of thermal activation, means that the effective activation energy is the sum of the elementary barriers associated with the motion of the objects that evolve cooperatively. In the Adam-Gibbs picture~\cite{AG}  these objects form a compact cluster, whereas in 
the frustration-limited domain picture~\cite{Gilles} and presumably also in the random first order transition approach near the thermodynamic singularity $T_{K}$~\cite{rfot,droplet} these objects form 
(possibly fractal) domain walls. In all cases, however,
the total effective activation energy is directly related to the number of correlated molecules, say $N_{\rm coop}$, through
\begin{equation}
\label{eq15:equation}
 \Delta(T)= N_{\rm coop}(T)^\theta \Delta_0,
\end{equation}
where $\Delta_0$ represents some elementary barrier that may or may not be temperature dependent, and $\theta$ is a certain exponent
accounting for possible renormalization effects.
As a result, $\tau_{\alpha} \propto \exp({N_{\rm coop}^\theta \Delta_0}/({k_{B}T}))$. Now, using the above definition of $N_{{\rm corr},T}$ in Eq.~(\ref{eq10:equation})
and considering for
illustration the simple picture in which $\theta=1$ and $\Delta_0$ 
is independent of temperature (as in the traditional Adam-Gibbs approach),
one obtains
\begin{equation}
\label{eq16:equation}
N_{{\rm corr},T} \sim \sqrt{\frac{k_{B}}{\Delta c_{p}}} \frac{\Delta_0}{k_{B}T}N_{\rm coop} 
\left[ 1 + \frac{\partial \log N_{\rm coop} }{\partial \log T} \right].
\end{equation}
Since $\Delta c_{p}$ is weakly temperature dependent, this
expression implies that $\Ncorr$ ($N_{{\rm corr},T}$ and, via Eq.~(\ref{eq11:equation}) $N_{{\rm corr},4}$ as well) increases as temperature decreases even when $N_{\rm coop}$ is temperature independent (and equal to one, for that matter, which is tantamount to an absence of ``cooperativity'').
This discussion  underlines that, contrary to the concept of ``dynamic heterogeneity'' that can be properly quantified, the notion of ``cooperativity'' in glassforming liquids lacks an operational definition that could then lead to concrete measurements of $N_{\rm coop}$.

We also point out that no meaningful relation seems to emerge between $N_{{\rm corr},T}$ and the stretching of the relaxation. Here, we do not refer to the way the stretching parameter enters in some expressions of $N_{{\rm corr},T}$ and to the fairly unintuitive behavior that follows (see Sec.~\ref{probe}). We rather consider the common view of the stretching as a phenomenon which is attributed to the spatial heterogeneity of the dynamics (and initially triggered the whole body of work on dynamic heterogeneities) and which is connected to the ``fragility'' of the glassforming materials~\cite{bohmer}. The main argument supporting the absence of link is the example of a simple system in which relaxation decays exponentially in time with a characteristic time that follows an Arrhenius temperature dependence. There is no stretching (and no ``fragility''), and yet, $N_{{\rm corr},T}$ increases as temperature decreases. It should be stressed again that our conclusions concerning $N_{{\rm corr},T}$ apply to $N_{{\rm corr},4}$ as well, since the two are intimately connected.

An additional illustration of the above points is provided is the case of the Fredrickson-Andersen (FA) models~\cite{FA}. FA models can be seen as defect models, where stochastic rules for the dynamical evolution of the defects are postulated. In the simplest form of the model (the ``one-spin facilitated FA model''), the relaxation of the system proceeds by the thermally activated diffusion of point defects throughout the system. The average relaxation time in the system behaves in an Arrhenius manner, while two-time correlation functions decay exponentially, at least in three spatial dimensions. Such a local relaxation process can hardly be said to be ``cooperative''. Yet, dynamics is very heterogeneous, in the sense that local dynamics is correlated on distances of the order of the typical distance between point defects, which can indeed become very large at low temperature. In this case, $\chi_4(t)$, $\chi_T$, and $\Ncorr$ all grow very strongly as temperature is decreased~\cite{steve,TWBBB,rob}.

For all the above reasons, we conclude on a  cautionary note: one should be
careful in comparing numerical and experimental values of $\Ncorr$ to
theoretical predictions made by theories which do not directly investigate
multi-point susceptibilities. This remark does not apply to mode-coupling
theory and kinetically constrained models, but does apply to most activation-based theories, such as the the Adam-Gibbs description~\cite{AG}, random-first order transition theory~\cite{rfot,droplet}, or the frustration-limited domain approach~\cite{Gilles}.

\begin{acknowledgments}
We thank T. Blochowicz for sharing the data analysis published in 
Refs.~\cite{refBlochowicz2006and2003}, R. Richert for the data 
in Ref.~\cite{OTP}, and F. Zamponi for discussions.
LPTL, LCP and LCVN are UMR 7600, 8000, and 5587 at the CNRS,
respectively. LB acknowledges financial support from the
Joint Theory Institute at the Argonne National Laboratory
and the University of Chicago. 
\end{acknowledgments}


\begin{thebibliography}{99}
\bibitem{footnote} Originally, this four-point time-dependent 
correlation function
was thought to be related to a dynamical susceptibility through the
fluctuation-dissipation theorem \cite{FP}. However, it turned out
that no such relation exists between the two quantities, correlation
function and dynamical susceptibility, despite the fact that their
behavior is similar in theoretical approaches such as the
mode-coupling theory -- see \cite{II}, Appendix A. For historical 
reason, the four-point function
$\chi_{4}(t)$ is now commonly called dynamic susceptibility,
although it does not directly represent a response of any kind. 
However, as shown in 
\cite{BBPRB}, the divergence, close to the glass transition, of 
the non-linear (cubic) response to an external field 
reflects that of $\chi_4$.
\end{thebibliography}
\end{document}